\documentclass[aps,prd,preprint,superscriptaddress,nofootinbib,floatfix]{revtex4-1} 

\usepackage{epsfig}
\usepackage{graphicx}
\usepackage[normalem]{ulem}
\usepackage{amssymb}
\usepackage{amsmath}
\usepackage{amsfonts}
\usepackage{latexsym}
\usepackage{verbatim}
\usepackage{mathtools}
\usepackage{setspace}
\usepackage{slashed}
\usepackage[all]{xypic}
\usepackage{psfrag}
\usepackage{comment}
\usepackage{array}
\usepackage{amssymb}
\usepackage{amsmath}
\usepackage{amsthm}
\usepackage{float}
\usepackage{amsfonts,wasysym,epsfig,verbatim,subfigure,bm,mathrsfs,lipsum}
\usepackage[colorlinks]{hyperref}
\usepackage{textcomp}

\begin{document}
	
\title{ Primordial black holes and stochastic gravitational wave background from inflation with a noncanonical spectator field }

\author{ Rong-Gen Cai }
\email{cairg@itp.ac.cn}
\affiliation{ CAS Key Laboratory of Theoretical Physics, Institute of Theoretical Physics, Chinese Academy of Sciences, Beijing 100190, China }
\affiliation{ School of Physical Sciences, University of Chinese Academy of Sciences, Beijing 100049, China } 

\author{ Chao Chen }
\email{cchao012@mail.ustc.edu.cn}
\affiliation{ Department of Astronomy, School of Physical Sciences, University of Science and Technology of China, Hefei, Anhui 230026, China }
\affiliation{Jockey Club Institute for Advanced Study, The Hong Kong University of Science and Technology, Hong Kong SAR, China }

\author{ Chengjie Fu }
\email[Corresponding author: ]{fucj@itp.ac.cn}
\affiliation{ CAS Key Laboratory of Theoretical Physics, Institute of Theoretical Physics, Chinese Academy of Sciences, Beijing 100190, China }
	
\begin{abstract}
We investigate the enhancement of the curvature perturbations in a single-field slow-roll inflation with a spectator scalar field kinetically coupled to the inflaton. The coupling term with a periodic function of inflaton triggers the exponential growth of the spectator field perturbations, which indirectly amplifies the curvature perturbations to produce a sizable abundance of primordial black holes (PBHs). This scenario is found to be insensitive to the inflationary background. We study two distinct populations of the stochastic gravitational wave background (SGWB) produced in this scenario, i.e., induced by the scalar perturbations during the inflationary era and the radiation-dominated era, respectively.
With the appropriate choices of parameter space, we consider two PBH mass windows of great interest. One is PBHs of masses $\mathcal{O}(10^{-12})M_\odot$ that can be a vital component of dark matter, and the predicted total energy spectrum of SGWB shows a unique profile and is detectable by LISA and Taiji. The other is PBHs of masses $\mathcal{O}(10)M_\odot$ which can provide consistent explanation for the LIGO-Virgo events. More interestingly, the predicted gravitational wave signal from the radiation-dominated era may account for the NANOGrav 12.5-yr results. 
\end{abstract}

\maketitle
	
\section{Introduction}
A region could collapse to a black hole by self-gravitation in the very early Universe as a result of initial overlarge density fluctuations \cite{Hawking:1971ei,Carr:1974nx,Carr:1975qj}. These primordial black holes (PBHs), in general, span a wide mass range from tens of micrograms to millions of solar masses due to their formation mechanism which is different from that of BHs who have stellar origins. In view of this promising feature, PBHs within different mass windows were expected to relate to a variety of cosmological and astronomical phenomena \cite{Belotsky:2014kca,Carr:2016drx,Carr:2020xqk,Bird:2016dcv,Sasaki:2016jop}. One of the major motivations to study PBHs is that they could serve as a reasonable candidate for the whole or an appreciable portion of dark matter (DM) which comprises $25\%$ of the critical density of our Universe \cite{Belotsky:2014kca,Carr:2016drx,Carr:2020xqk}. In particular, the suggestion that the gravitational waves (GWs) were generated by the coalescence of PBH binaries has attracted much attention in recent years with the LIGO-Virgo detection \cite{Abbott:2016blz,Abbott:2016nmj,Abbott:2017vtc,Mukherjee:2021ags,Mukherjee:2021itf}. Shortly after the first detection of the GW event, GW150914 \cite{Abbott:2016blz}, Bird \emph{et al.} \cite{Bird:2016dcv} found that the merger rate of PBHs with masses of $\sim 30M_\odot$ falls within the range of $2$--$53$~Gpc$^{-3}$~yr$^{-1}$ inferred from GW150914, and Sasaki \emph{et al.} \cite{Sasaki:2016jop} argued that the expected PBH merger rate can be made compatible if PBHs constitute $\mathcal{O}(1)$\textperthousand\,of DM.

The induced GWs associated with PBH formation have attracted great attention in recent years. According to second-order cosmological perturbation theory \cite{Matarrese:1997ay,Noh:2004bc,Ananda:2006af,Baumann:2007zm,Wang:2017krj,Domenech:2017ems}, the scalar perturbations would provide anisotropic stress via the nonlinear couplings and result in the generation of GWs. Hence, the overlarge primordial curvature perturbations for PBH formation should induce sizeable GW signals after their reentry into the Hubble horizon during the radiation-dominated era \cite{Saito:2008jc,Saito:2009jt,Bugaev:2010bb,Garcia-Bellido:2017aan,Espinosa:2018eve,Kohri:2018awv,Bartolo:2018rku,Cai:2018dig,Cai:2019jah,Fu:2019vqc,Cai:2019amo,Pi:2019ihn,Cai:2019elf,Inomata:2020cck,Pi:2020otn,Braglia:2020eai,Braglia:2020taf,Peng:2021zon}. This strong correlation between PBHs and the concomitant GW signals could be a promising approach to detecting PBHs in the upcoming GW experiments, such as space-based projects LISA \cite{LISA:2017pwj} and Taiji \cite{Ruan:2018tsw}. In addition, the North American Nanohertz Observatory for Gravitational Waves (NANOGrav) \cite{Arzoumanian:2020vkk} recently has reported strong evidence of a stochastic common-spectrum process across pulsars from analyzing 12.5-yr pulsar timing array data, which might be interpreted as the signal of a stochastic GW background (SGWB). Although NANOGrav did not find the significant ``quadrupole" or ``Hellings-Downs (HD)" correlation which would have been a  smoking gun for a SGWB \cite{Hellings:1983fr}, it is worthwhile to study the potential implications of the NANOGrav signal in terms of SGWB.
To date, there are already many works discussing the interpretation for a NANOGrav signal, such as from the cosmic string \cite{Blasi:2020mfx, Ellis:2020ena, Buchmuller:2020lbh, Samanta:2020cdk},  phase transition \cite{Nakai:2020oit,Neronov:2020qrl,Bian:2021lmz,Abe:2020sqb} , domain wall~\cite{Liu:2020mru}, and also as  inflationary GWs \cite{Vagnozzi:2020gtf,Kuroyanagi:2020sfw} and SGWB related to PBH formation \cite{Vaskonen:2020lbd, DeLuca:2020agl, Kohri:2020qqd, Inomata:2020xad, Yi:2021lxc, Sugiyama:2020roc, Domenech:2020ers}. 

Currently, the most popular scenario for PBH formation within the inflationary Universe is to generate large-amplitude primordial curvature perturbations during inflation. In this sense, PBHs could be used to detect the physics of the cosmic inflation. The key challenge of this scenario is to enhance the primordial curvature perturbations on small scales, while one has to ensure the nearly scale-independent property of the large-scale perturbations that was confirmed by cosmic microwave background (CMB) observations \cite{Akrami:2018odb} and large-scale structure (LSS) surveys \cite{Tegmark:2003uf} with high precision. Up to now, there are a tremendous amount of the mechanisms for enhancing small-scale curvature perturbations, such as the ultra-slow-roll inflation \cite{Garcia-Bellido:2017mdw,Germani:2017bcs,Motohashi:2017kbs,Ezquiaga:2017fvi,Gong:2017qlj,Ballesteros:2017fsr,Dalianis:2018frf,Gao:2018pvq,Drees:2019xpp,Xu:2019bdp,Fu:2019ttf,Lin:2020goi,Fu:2020lob,Yi:2020cut}, the small sound speed \cite{Ballesteros:2018wlw,Kamenshchik:2018sig}, the parametric resonance \cite{Cai:2018tuh,Chen:2019zza, Chen:2020uhe,Cai:2019bmk,Zhou:2020kkf,Liu:2021rgq}, the modified dispersion relation for scalar perturbations~\cite{Ashoorioon:2019xqc}, etc.

Among the existing mechanisms for enhancing curvature perturbations, the parametric resonance that triggers the exponential growth of the scalar field perturbations leads, in general, to a phenomenology that is much richer than that of the other mechanisms. In particular, the sound speed resonance predicts a unique energy spectrum of GWs with a double-peak profile due to the contribution of the induced GWs during inflation \cite{Cai:2019jah}. Moreover, there has been an increasing interest in the possibility of enhancing curvature perturbations through the resonance of the spectator fields which play no role in driving inflation. A common scenario is to consider the model of an axionlike inflaton field interacting with a gauge field via Chern-Simons type coupling \cite{Cheng:2018yyr,Ozsoy:2020kat}, in which the temporary fast-roll of inflaton around the clifflike regions of potential triggers the production of one helicity state of gauge fields.  References. \cite{Cook:2011hg, Goolsby-Cole:2017hod} studied the possibility of amplifying GWs from the coupling between the inflaton and spectator field. Furthermore, there are some works focusing on amplification of primordial perturbations through the small sound speed of spectator fields, e.g., Refs. \cite{Biagetti:2013kwa,Biagetti:2014asa,Fujita:2014oba}. In this work, we consider the enhancement of the curvature perturbations as the result of the kinetic coupling between the  inflaton and a spectator scalar field. This kinetic coupling makes the energy transfer between the inflaton and the spectator field more efficient, and the inflaton fluctuations would be amplified accompanying with the resonance of the spectator field. In contrast to the discussions in the  existing literatures, the resonating spectator field can lead to the exponential growth of the curvature perturbations even at the linear level. The resulting PBH formation and induced SGWB in this model are studied numerically.

The organization of this paper is as follows. In the next section, we describe the model of inflation with a spectator scalar field and discuss the possibility of enhancing the curvature perturbations through this sector. In Sec. \ref{PBH_IGW}, we outline the estimation of PBH abundance as well as the energy spectra  of GWs from the inflationary era and the radiation-dominated era, respectively. In Sec. \ref{results}, we present the results of our numerical calculations for the curvature perturbations, PBHs and SGWB. Finally, the conclusions are given in Sec. \ref{conclusions}.

\section{Model and linearized dynamics}
\label{Model}

\subsection{Basic equations}
Our model consists of two scalar fields minimally coupled to gravity, one of which has a noncanonical kinetic term, and is specified by the following action:
\begin{align}\label{action}
\mathcal{S}=\int d^4x \sqrt{-g}\left[\frac{M_{\mathrm{p}}^2}{2} R - \frac{1}{2}(\nabla\phi)^2 - \frac{e^{2b(\phi)}}{2}(\nabla\chi)^2 - V(\phi,\chi) \right]\,,
\end{align}
where $M_{\mathrm{p}}=2.4\times10^{18}$ GeV is the reduced Planck mass. Such an action is motivated by many of
generalized Einstein theories after making use of a conformal transformation \cite{Starobinsky:2001xq}. In this model, we consider that the inflation is driven by the canonical field $\phi$, in the presence of a noncanonical spectator field $\chi$.
Due to the presence of the function $b(\phi)$, through which the two fields can interact, we consider the following decoupled potential where the spectator field has a simple quadratic potential:
\begin{align}
V(\phi,\chi) = V(\phi) + \frac{1}{2}m_\chi^2 \chi^2\,
\end{align}
with an arbitrary inflation potential favored by the current observational data.

Throughout this paper, we work with the spatially flat Friedmann-Lema\^{\i}tre-Robertson-Walker metric in the conformal Newtonian gauge. Ignoring the anisotropic stress, the perturbed metric can then read
\begin{align}\label{line_element}
ds^2 = -(1+2\Psi)dt^2 + a(t)^2 \left[(1-2\Psi)\delta_{ij} + \frac{1}{2}h_{ij} \right] dx^idx^j\,,
\end{align}
where the first-order scalar metric perturbations are described by the single variable $\Psi$, and $h_{ij}$ represents the second-order transverse-traceless tensor perturbations. The reason that we choose such a gauge for the perturbed metric is that one of the main focuses of this paper is on SGWB induced by the scalar perturbations, whose calculations and physical interpretations are both suitable to be performed in the conformal Newtonian gauge \cite{Domenech:2020xin}.

During inflation, we split the scalar fields into a homogeneous background and a first-order perturbation,
\begin{align}
\phi(t,\textbf{x}) = \phi_0(t) + \delta\phi(t,\textbf{x})\,, \qquad	\chi(t,\textbf{x}) = \chi_0(t) + \delta\chi(t,\textbf{x})\,.
\end{align}
The dynamics of the homogeneous scalar fields and the scale factor can be solved from the action \eqref{action}
\begin{align}
\label{Friedamn}-2\dot H  = M^{-2}_{\rm{p}} \left( \dot{\phi}_{0}^2 +   e^{2b} \dot{\chi}_{0}^2 \right) \,, \\
\ddot{\phi}_0  +  3H\dot{\phi}_0 - b_{,\phi} e^{2b} \dot{\chi}_{0}^2 + V_{,\phi}=0\,,\\
\ddot{\chi}_0 +  \left(3H + 2 b_{,\phi} \dot{\phi}_{0} \right)\dot{\chi}_{0} +  e^{-2b} m_\chi^2 \chi_0=0\,,
\end{align}
where $H\equiv \dot a/a$ represents the Hubble parameter. The first-order perturbed Einstein and Klein-Gordon equations are given by
\begin{align}
&3H \dot\Psi + \left(\frac{k^2}{a^2}+3H^2 \right) \Psi \nonumber  \\
& \qquad  = \frac{1}{2M_{\rm{p}}^{2}}\left[\left(\dot{\phi}_{0}^2 + e^{2b} \dot{\chi}_{0}^2 \right) \Psi  - \dot\phi_0 \delta\dot\phi - e^{2b} \dot{\chi}_0\delta\dot\chi - \left(b_{,\phi} e^{2b}\dot{\chi}_{0}^2  + V_{,\phi} \right)\delta\phi  - m_\chi^2 \chi_0\delta\chi \right]\,, \\
\label{Psi}&\dot \Psi + H\Psi = \frac{1}{2M_{\rm{p}}^{2}}\left( \dot\phi_0\delta\phi  +   e^{2b} \dot{\chi}_{0} \delta\chi \right)  \,,\\
\label{delta_phi}&\delta\ddot\phi  +3H \delta\dot\phi - \frac{\Delta}{a^2}\delta\phi +  \left[V_{,\phi\phi}- \left(b_{,\phi\phi} + 2b_{,\phi}^2 \right)e^{2b} \dot{\chi}_{0}^2\right]\delta\phi  - 2 b_{,\phi} e^{2b}\dot{\chi}_{0}\delta\dot\chi  = -2V_{,\phi}\Psi  + 4 \dot\phi_0\dot\Psi \,, \\
\label{EoM_delta_chi} & \delta\ddot\chi   + \left(3H + 2 b_{,\phi} \dot{\phi}_{0} \right) \delta\dot\chi - \frac{\Delta}{a^2}\delta\chi +  e^{-2b}m_\chi^2 \delta\chi + 2b_{,\phi}\dot{\chi}_{0}\delta\dot\phi  + \left( 2b_{,\phi} e^{-2b} m_\chi^2  \chi_0 + 2b_{,\phi\phi}\dot{\phi}_{0}\dot{\chi}_{0}\right)\delta\phi \nonumber \\
& \qquad  =  -2 e^{-2b}m_\chi^2\chi_0\Psi  + 4 \dot\chi_0\dot\Psi \,.
\end{align}

\subsection{ Amplification of curvature perturbations from the resonance of $\chi$ field}
As the standard quantization process, we promote the field $\chi$ to quantum operator $\hat\chi$, and the Fourier components $\delta\chi_{\bf k}$ defined by $\delta\chi(t,{\bf x}) = (2\pi)^{-3/2}\int d^3k \delta\chi_{\bf k}(t) e^{i {\bf k} \cdot {\bf x}}$ are expressed via the following decomposition,
\begin{align}
 \delta\hat\chi_{\bf{k}}(t)=\delta\chi_k(t)\hat a_{\bf{k}}+\delta\chi_{-k}^\ast(t)\hat a^\dagger_{-\bf{k}}\,,
\end{align}
where the creation and annihilation operators $\hat a^\dagger_{\bf{k}}$ and $\hat a_{\bf{k}}$ satisfy the canonical commutation relation $[\hat a_{\bf{k}},\hat a^\dagger_{\bf{k}^\prime}] = \delta({\bf k} - {\bf k}^\prime)$. It is useful to make a similar operation for inflaton fluctuation $\delta\phi$. Using Eq. (\ref{EoM_delta_chi}) with the $\chi_0$ background set to zero, the mode function $\delta\chi_k$ obeys 
\begin{align}\label{EoM_mode_function}
\delta\ddot\chi_k   + \left(3H + 2 b_{,\phi} \dot{\phi}_{0} \right) \delta\dot\chi_k +  \left( \frac{k^2}{a^2} + e^{-2b}  m_\chi^2 \right) \delta\chi_k = 0\,.
\end{align}
Through transforming $\delta\chi_k$ as
\begin{align}
	X_k \equiv a^{3/2}e^b\delta\chi_k\,,
\end{align}
Eq. (\ref{EoM_mode_function}) can be reexpressed as 
\begin{align}\label{EoM_X}
	\ddot X_k + \left(\frac{k^2}{a^2} + m_{\chi,\mathrm{eff}}^2\right)X_k=0
\end{align}
with the effective mass term
\begin{align}\label{m_eff}
	m_{\chi,\mathrm{eff}}^2 = e^{-2b}m_\chi^2 -\frac{9}{4}H^2-\frac{3}{2}\dot H - 3Hb_{,\phi}\dot\phi_0 - b_{,\phi}^2\dot\phi_0^2 - b_{,\phi\phi}\dot\phi_0^2 - b_{,\phi}\ddot\phi_0\,.
\end{align}

In this paper, we phenomenologically consider that $b(\phi)$ is a periodic function of $\phi$ [see Ref. \cite{vandeBruck:2014ata} that studied the linear and quadratic form of $b(\phi)$], and take the following form as a typical representative:
\begin{align}
b(\phi)=\frac{\xi}{2}\cos\left(\frac{\phi}{\phi_c}\right)\Theta(\phi_s-\phi_0)\Theta(\phi_0-\phi_e)\,,
\end{align}
where $\xi$ is a dimensionless parameter and $\phi_c$ is the characteristic field value. Here $\Theta$ is the Heaviside function, and this profile describes that the two fields nongravitationally interact with each other just during the period when the inflaton rolls from $\phi_0=\phi_s$ to $\phi_0=\phi_e$. We only focus on this period in the subsequent discussion. 
By appropriately choosing parameters $\xi$ and $\phi_c$, which satisfy $\xi<1$ and $|\dot\phi_0|/\phi_c \gg H$, the $b_{,\phi\phi}$ term may dominate the square of the effective mass given in Eq. (\ref{m_eff}). Then, Eq. (\ref{EoM_X}) can be simplified as 
\begin{align}\label{EoM_X_simple}
\ddot X_k + \left[\frac{k^2}{a^2} + \frac{\xi\dot\phi_0^2}{2\phi_c^2}\cos\left(\frac{\phi_0}{\phi_c}\right)\right]X_k=0\;.
\end{align}
Since the inflaton is slowly rolling during inflation, the evolution of the $\phi$ background can be simply described as $\phi_0 = c + \dot\phi_0 t$ with $c$ being a constant during a small field excursion $\Delta\phi \equiv |\phi_s-\phi_e| \ll M_{\rm p}$. Thus, after introducing a dimensionless time parameter $2z = (c + \dot\phi_0 t )/\phi_c + \pi$, Eq. (\ref{EoM_X_simple}) can be cast in the form of the so-called Mathieu equation,
\begin{align}\label{Mathieu}
\frac{d^2X_k}{dz^2}+\left[A_k - 2q\cos(2z)\right]X_k=0\;,
\end{align}
where $A_k =4k^2\phi_c^2/(a^2\dot\phi_0^2)$ and $q=\xi$. In particular, Floquet theory states that $X_k$ will grow exponentially, known as the parametric resonance, when ($A_k, q$) fall in an instability band of the Mathieu equation \eqref{Mathieu}. For $q<1$, namely, $\xi<1$ chosen in our paper, the most enhanced instability band is located in the narrow region with $ |A_k -1| < q$, and obviously each mode will pass through the instability band as a result of the expansion of the Universe. However, since Eq. \eqref{Mathieu} is valid during the period from $\phi_0=\phi_s$ to $\phi_0=\phi_e$, only the perturbation $\delta\chi_k$ (or $X_k$), with a $k$ mode that is redshifted to the instability band during this period, will grow exponentially. Thus, the resonant amplification of $\delta\chi_k$ only appears in the specific scales.
In addition, since $|\dot\phi_0|/\phi_c \gg H$ the resonance of each mode occurs deep inside the Hubble horizon.

For the case of two scalar fields, the definition of comoving curvature perturbation is given by \cite{Gordon:2000hv}
\begin{align}
\mathcal{R} \equiv \Psi + H \frac{ \dot\phi_0\delta\phi + e^{2b} \dot{\chi}_{0}\delta\chi }{\dot{\phi}_{0}^2 + e^{2b} \dot{\chi}_{0}^2}\,.
\end{align}
By using Eqs. (\ref{Friedamn}) and (\ref{Psi}), the comoving curvature perturbation can also be given in terms of the metric perturbations in the conformal Newtonian gauge,
\begin{align}
\mathcal{R} = \Psi - \frac{H}{\dot H} \left( \dot \Psi + H\Psi \right) \,.
\end{align}
In the above  discussions, we have ignored the $\chi$ background. According to Eq. (\ref{Psi}), the metric perturbations couple only to the inflaton perturbations in the absence of a $\chi$ background, at first-order level. Besides, it is evident, as seen from Eq. (\ref{delta_phi}), that the evolution of $\delta\phi$ does not feel the appearance of $\delta\chi$ without a $\chi$ background. As a result, the $\chi$ field perturbations resonantly amplified do not trigger the enhancement of curvature perturbations at the linear level. In the present paper, we consider the spectator field $\chi$ with a homogeneous background $\chi_0$, which gives a negligible contribution to the energy density during inflation. Moreover, we expect that the $\chi$ field background and perturbations decay after the end of inflation, in view of the $\chi$ field acting as a spectator, rather than a curvaton. Therefore, we consider a massive $\chi$ field with $m_\chi \sim \mathcal{O}\left(H\right)$ in the present paper, and the modes of $\chi$ field perturbations will decrease after crossing the Hubble horizon. 
In addition, it should be note that the entropy perturbation also quickly decreases as the decay of $\delta\chi$, since the entropy perturbation is due solely to $\delta\chi$ on large scales when the background $\chi_0$ vanishes. So we can safely ignore the remaining entropy perturbation in our model, which is consistent with the CMB observations.
To sum up, the resonant modes $\delta\chi_k$ induce the exponential growth of the corresponding modes $\delta\phi_k$ through nongravitational interaction inside the horizon and, in turn, enhance indirectly the curvature perturbations at some certain scales.

\section{Primordial black holes and stochastic gravitational wave background}
\label{PBH_IGW}
In this section, we give a brief review of PBH formation originated from the primordial curvature perturbations and of the SGWB produced both during the inflationary era and the radiation-dominated era.

\subsection{Primordial black holes}
When primordial curvature perturbations with large amplitudes at a certain scale reenter the Hubble horizon at the radiation-dominated epoch, the resulting overdense regions will collapse to form PBHs, whose masses $M$ can be related to the comoving wave number $k$ of the sourced primordial curvature perturbations as \cite{Ando:2018qdb}
\begin{align}
M = \gamma \frac{4\pi M_{\rm p}^2 }{H} \bigg|_{k=aH}  \simeq M_\odot \left(\frac{\gamma}{0.2}\right)\left(\frac{g_{\ast,{\rm form}}}{10.75}\right)^{-1/6}\left(\frac{k}{1.9\times10^6\;{\rm Mpc}^{-1}}\right)^{-2}\,,
\end{align}
where $g_{\ast,{\rm form}}$ is the effective number of the relativistic degree of freedom at PBH formation. We adopt $g_{\ast,{\rm form}}=106.75$ as a fiducial value in this paper. Here, the correction factor $\gamma$, representing the mass efficiency of collapse, can be evaluated as $\gamma \simeq 0.2$ through a simple analytical calculation \cite{Carr:1975qj}. If the perturbations follow the Gaussian probability distribution function, the mass fraction $\beta$ of PBHs formed is approximately given by \cite{Ando:2018qdb}
\begin{align}
\beta(M) = \gamma \int_{\delta_c} \frac{d\delta}{\sqrt{2\pi\sigma^2(M)}}e^{-\frac{\delta^2}{2\sigma^2(M)}}\simeq \frac{\gamma}{\sqrt{2\pi}\delta_c/\sigma(M)} e^{-\frac{\delta_c^2}{2\sigma^2(M)}}\,,
\end{align}
where $\delta_c\simeq 0.45$ is the threshold for PBH formation \cite{Musco:2004ak}.Here,  $\sigma^2(M)$ represents the variance of density contrast given by \cite{Blais:2002gw,Josan:2009qn}
\begin{align}
\sigma^2(M(k)) = \int d\ln q \tilde{W}(qk^{-1})\frac{16}{81}\left(qk^{-1}\right)^4T^2(q,\eta=k^{-1})\mathcal{P_R}(q)\,,
\end{align}
where $\mathcal{P_R}$ is the power spectrum of $\mathcal{R}$ in the superhorizon limit and $T(q,\eta)$ is the transfer function at the radiation-dominated phase defined as
\begin{align}
T(q,\eta) = 3 \frac{\sin(q\eta/\sqrt{3})-(q\eta/\sqrt{3})\cos(q\eta/\sqrt{3})}{(q\eta/\sqrt{3})^3}\,.
\end{align}
Here, $\tilde{W}$ is the windows function,  whose different choices will affect the corresponding relation between the PBH abundance and the required power spectrum. For instance, compared to the Gaussian and $k$-space top-hat ones, the required amplitude of the power spectrum with the real-space top-hat window function is the smallest one for the same abundance of PBHs \cite{Ando:2018qdb}. In this paper, we adopt the the real-space top-hat window function given as follows:
\begin{align}
\tilde{W}(x) = 3\left( \frac{\sin(x)-x\cos(x)}{x^3} \right)\,.
\end{align}
The fraction of PBHs against the total DM at present is given by \cite{Sasaki:2018dmp}
\begin{align}
f_{\rm PBH} (M) \equiv \frac{\Omega_{\rm PBH}(M)}{\Omega_{\rm DM}} = 2.7\times 10^8 \left(\frac{\gamma}{0.2}\right)^{1/2}\left(\frac{g_{\ast,{\rm form}}}{10.75}\right)^{-1/4}\left(\frac{M	}{M_\odot}\right)^{-1/2}\beta(M)\,.
\end{align}

\subsection{Stochastic gravitational wave background}
In the model introduced in this work, there are two distinct populations of the induced GW background: one originates from the amplified scalar field perturbations during inflation, and the other is associated with the enhanced primordial curvature perturbations that reenter the horizon to form PBHs in the radiation-dominated era. Next, we derive the basic formulas of the induced GWs from the inflationary era and the radiation-dominated era, respectively.

Under the conformal time $\tau\equiv \int^tdt/a$, the tensor perturbations, $h_{ij}$, obey the following equation of motion:
\begin{align}\label{EoM_hij}
h_{ij}^{\prime\prime}(\tau,\textbf{x}) + 2\mathcal{H}h_{ij}^{\prime}(\tau,\textbf{x}) - \nabla^2h_{ij}(\tau,\textbf{x}) = 4\mathcal{\hat T}^{lm}_{ij}S_{lm}(\tau,\textbf{x})\,,
\end{align}
where a prime denotes the derivative with respect to $\tau$, $\mathcal{H}\equiv a^\prime/a$ the conformal Hubble parameter, and $\mathcal{\hat T}^{lm}_{ij}$ the transverse-traceless projection operator. We expand the tensor perturbation $h_{ij}$ and the source $\mathcal{\hat T}^{lm}_{ij}S_{lm}$ in Fourier space, respectively, as 
\begin{align}\label{hij_source_Fourier}
h_{ij}(\tau,\textbf{x})&= \sum\limits_{\lambda=+,\times}\int \frac{d^3\bf{k}}{(2\pi)^{3/2}}e^{i\bf{k}\cdot \bf{x}} e^\lambda_{ij}(\textbf{k})h^\lambda_{\textbf{k}}(\tau)\,, \nonumber \\
\mathcal{\hat T}^{lm}_{ij}S_{lm}(\tau,\textbf{x})&= \sum\limits_{\lambda=+,\times}\int \frac{d^3\bf{k}}{(2\pi)^{3/2}}e^{i\bf{k}\cdot \bf{x}}e^\lambda_{ij}(\textbf{k})e^{\lambda, lm}(\textbf{k})S_{lm}(\tau,\textbf{k})\,,
\end{align}
where $S_{lm}(\tau,\textbf{k})$ is the Fourier transform of $S_{lm}(\tau,\textbf{x})$, and the index $\lambda=+, \times$ denotes two polarization states of GWs. Here, the polarization tensors $e^\lambda_{ij}(\textbf{k})$ are transverse traceless and satisfy the conditions $k^ie^\lambda_{ij}(\textbf{k}) =0$, $e^{\lambda}_{ij}(\textbf{k})e^{\lambda^\prime ,ij}(\textbf{k})=\delta^{\lambda\lambda^\prime}$, and $e^{\lambda}_{ij}(-\textbf{k})=e^{\lambda}_{ij}(\textbf{k})$. Equations \eqref{EoM_hij} and \eqref{hij_source_Fourier} can be combined into the equation
\begin{align}\label{EoM_hij_Fourier}
h_{\textbf{k}}^{\lambda\prime\prime}(\tau) + 2\mathcal{H}h_{\textbf{k}}^{\lambda\prime}(\tau) + k^2h_{\textbf{k}}^{\lambda}(\tau) = S^\lambda_\textbf{k}(\tau)
\end{align}
with
\begin{align}\label{S_k}
S^\lambda_\textbf{k}(\tau) = 4e^{\lambda,ij}(\textbf{k})S_{ij}(\tau,\textbf{k}) = 4 \int \frac{d^3\bf{x}}{(2\pi)^{3/2}}e^{-i\bf{k}\cdot \bf{x}}e^{\lambda,ij} S_{ij}(\tau,\textbf{x})\,.
\end{align}
Through the Green's function method, Eq. \eqref{EoM_hij_Fourier} is solved by
\begin{align}\label{solution_h_k}
	h_{\textbf{k}}^{\lambda}(\tau) = \int^\tau d\tau^\prime g_{k}(\tau,\tau^\prime)S^\lambda_\textbf{k}(\tau^\prime)\,,
\end{align}
where $g_{k}(\tau,\tau^\prime)$ is the Green's function. The power spectrum for a single polarization of tensor perturbations is defined as
\begin{align}
\langle \hat h_{\textbf{k}}^{\lambda}(\tau) \hat h_{\textbf{k}^\prime}^{\lambda^\prime}(\tau)\rangle = \delta^{\lambda\lambda^\prime}\delta^{(3)}(\textbf{k}+\textbf{k}^\prime)\frac{2\pi^2}{k^3}\mathcal{P}^\lambda_h(\tau,k)\,.
\end{align}

\subsubsection{Inflationary era}
During inflation, $S_{ij}(\tau,\textbf{x})$ given in Eq. \eqref{EoM_hij} has the following form:
\begin{align}\label{S_ij_inf}
S_{ij}(\tau,\textbf{x}) = -4\Psi\partial_i\partial_j\Psi - 2\partial_i\Psi\partial_j\Psi + M^{-2}_{\rm{p}}\left(\partial_i\delta\phi\partial_j\delta\phi + e^{2b} \partial_i\delta\chi\partial_j\delta\chi \right)\,,
\end{align}
Since the coupling between $\Psi$ and the scalar fields is slow-rolling  suppressed, $\Psi$ plays no role in the emission of GWs, and therefore, we choose to neglect the $\Psi$ terms in Eq. \eqref{S_ij_inf}. Then, we have
\begin{align}\label{S_k}
S^\lambda_\textbf{k}(\tau)= \frac{4}{M_{\rm p}^2}\int \frac{d^3\bf{p}}{(2\pi)^{3/2}} e^{\lambda, ij}(\textbf{k}) p_i p_j \left( \delta\phi_{\bf{p}}(\tau) \delta\phi_{\textbf{k}-\textbf{p}}(\tau)+ e^{2b}\delta\chi_{\bf{p}}(\tau) \delta\chi_{\textbf{k}-\textbf{p}}(\tau) \right)\,.
\end{align}
Proceeding in the standard fashion, the power spectrum of tensor perturbations, $\mathcal{P}_h = \sum_{\lambda=+,\times} \mathcal{P}_h^\lambda$, is given by
\begin{align}
\mathcal{P}_h(\tau,k) =\frac{2k^3}{\pi^4M_{\mathrm{p}}^4} \int^{\infty}_{0} dp p^6 \int^{1}_{-1} & d\cos\theta \sin^4 \theta \left(\left|\int^\tau d\tau^\prime g_{k}(\tau,\tau^\prime)\delta\phi_p(\tau^\prime)\delta\phi_{|\textbf{k}-\textbf{p}|}(\tau^\prime)\right|^2 \right.\nonumber\\
&\left.+\left|\int^\tau d\tau^\prime g_{k}(\tau,\tau^\prime)e^{2b}\delta\chi_p(\tau^\prime)\delta\chi_{|\textbf{k}-\textbf{p}|}(\tau^\prime)\right|^2 \right)\,,
\end{align}
where the Green's function $g_{k}(\tau,\tau^\prime)$ during inflation with $a(\tau)\simeq -1/H\tau$ reads \cite{Biagetti:2013kwa}
\begin{align}
g_{k}(\tau,\tau^\prime)=\frac{1}{k^3\tau^{\prime2}}[-k(\tau-\tau^\prime)\cos(k\tau-k\tau^\prime)+(1+k^2\tau\tau^\prime)\sin(k\tau-k\tau^\prime)]\Theta(\tau-\tau^\prime)\,.
\end{align}
When the tensor modes sourced by the scalar fields during inflation reenter the Hubble horizon, they become a SGWB and evolve into the present. For the waves with frequency $f>10^{-10}\;{\rm Hz}$, the current energy spectrum of SGWB is related to the inflationary power spectrum of the tensor perturbations via \cite{Zhao:2006mm}
\begin{align}
\Omega_{\mathrm{GW},0}^{\rm (inf)}(k) = 2.7\times 10^{-7}\mathcal{P}_h(\tau_{\rm end},k)\,,
\end{align}
where $\tau_{\rm end}$ denotes the time at the end of inflation.

\subsubsection{Radiation-dominated era}
In the radiation-dominated era, $S_{ij}(\tau,\textbf{x})$ is given by \cite{Ananda:2006af,Baumann:2007zm}
\begin{align}
S_{ij}(\tau,\textbf{x})=-4\Psi\partial_i\partial_j\Psi-2\partial_i\Psi\partial_j\Psi+\frac{1}{\mathcal{H}^2}\partial_i(\mathcal{H}\Psi+\Psi^\prime)\partial_j(\mathcal{H}\Psi+\Psi^\prime)\,.
\end{align}
The evolution of the mode function $\Psi_k$ may be described as \cite{Ananda:2006af}
\begin{align}
\Psi_k(\tau) = \frac{\psi_k}{(k\tau)^3}\left[ \frac{k\tau}{\sqrt{3}}\cos\left(\frac{k\tau}{\sqrt{3}}\right) - \sin\left(\frac{k\tau}{\sqrt{3}}\right) \right]\,,
\end{align}
where $\psi_k$ is related to the power spectrum of primordial curvature perturbations as
\begin{align}
\psi_k^2 = \frac{216\pi^2}{k^3}\mathcal{P_R}(k)\,.
\end{align}
The density parameter of the induced GWs per logarithmic interval of $k$ is given by
\begin{align}
\Omega_{\rm GW}(\tau,k) = \frac{1}{48}\left(\frac{k}{a(\tau)H(\tau)}\right)^2\overline{\mathcal{P}_h(\tau,k)},
\end{align}
where the overline denotes the oscillation average and the power spectrum for GWs is written as
\begin{align}
\mathcal{P}_h(\tau,k) = 8\int^\infty_0 dv \int^{|1+v|}_{|1-v|}du \left( \frac{4v^2-(1+v^2-u^2)^2}{4uv}\right)^2I^2(v,u,k\tau)\mathcal{P}_\mathcal{R}(ku)\mathcal{P}_\mathcal{R}(kv)\,.
\end{align}
The oscillation average of the function $I^2$ in the subhorizon limit ($x=k\tau\rightarrow \infty$) can be evaluated as \cite{Kohri:2018awv}
\begin{align}
\overline{I^2(v,u,x\rightarrow \infty)} =& \frac{1}{2}\left( \frac{3(u^2+v^2-3)}{4u^3v^3x}\right)^2 \bigg(\bigg(-4uv+(u^2+v^2-3) \ln\left| \frac{3-(u+v)^2}{3-(u-v)^2}\right| \bigg)^2  \nonumber \\
& + \pi^2(u^2+v^2-3)^2\Theta(v+u-\sqrt{3})\bigg)\,.
\end{align}
We set $\tau=\tau_c$ to represent the time when the induced GWs start to evolve without any source, and the relation between the energy spectrum at present and that at $\tau_c$ is derived as \cite{Ando:2018qdb}
\begin{align}
\Omega_{\rm GW,0}^{\rm (rad)} = 0.83\left(\frac{g_{\ast,c}}{10.75}\right)^{-1/3}\Omega_{\rm r,0}\Omega_{\rm GW}(\tau_c,k)\,,
\end{align}
where we simply take $g_{\ast,c} = g_{\ast,{}\rm form}$ and $\Omega_{\rm r,0}$ represents the current density parameter of radiation.

\begin{table}
	\caption{Three sets of parameters given in the coupling function $b(\phi)$}
	\begin{tabular}{>{\centering}p{1.cm}>{\centering}p{2cm}>{\centering}p{2cm}>{\centering}p{2cm}>{\centering}p{2cm}}
		\hline
		\hline
		Set & $\xi$ & $\phi_c/M_{\mathrm{p}}$ & $\phi_s/M_{\mathrm{p}}$ & $\phi_e/M_{\mathrm{p}}$  \tabularnewline
		\hline
		
		1 & $0.7724$ & $5\times10^{-4}$ & $4.70$ & $4.66$ \tabularnewline
		
		2 & $0.773$ & $5\times10^{-4}$ & $4.75$ & $4.65$  \tabularnewline
		
		3 & $0.808$ & $5\times10^{-4}$ & $5.16$ & $5.05$ \tabularnewline
		\hline
	\end{tabular}
	\label{table1}
\end{table}

\begin{figure}
	\centering
	\includegraphics[width=0.8\textwidth ]{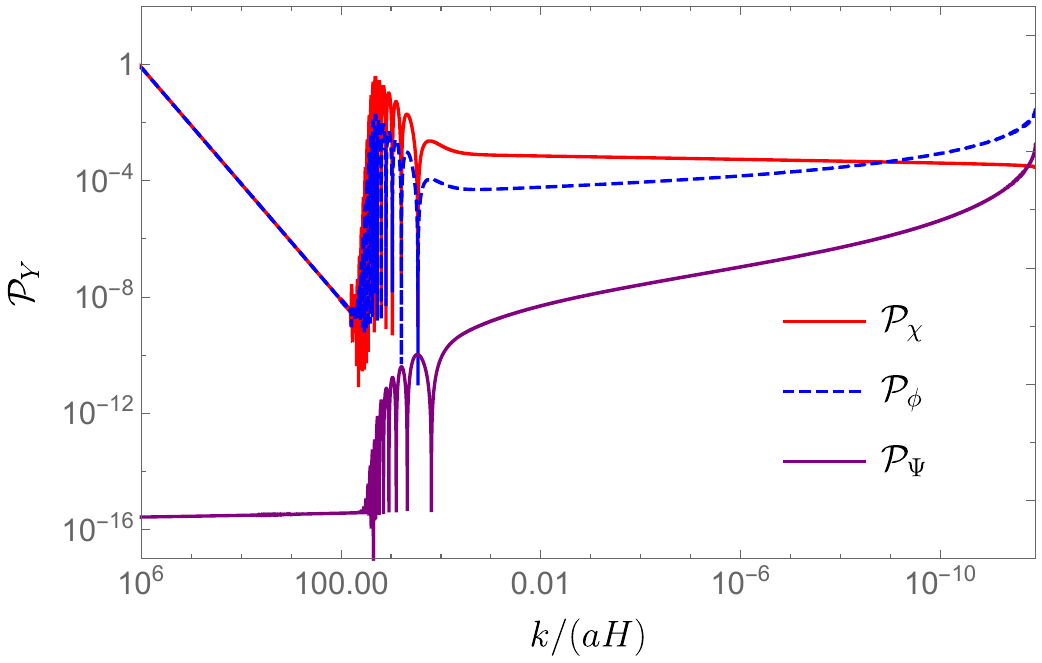}
	\caption{\label{fig1} The evolution of the power spectrum $\mathcal{P}_Y(k)\equiv k^3/(2\pi^2)|Y_k|^2$, for $Y_k=\delta\chi_k$, $\delta\phi_k$ and $\Psi_k$, as a function of $k/(aH)$ in the case of using Set 1. This particular comoving wave number $k$ is redshifted to the resonance band during the period when the inflaton goes through $\phi_s$ to $\phi_e$.}
\end{figure}

\section{numerical results}
\label{results}
According to the above discussions, there are not any theoretical restrictions on the inflaton potential, which implies our model is insensitive to the inflationary background. For convenience, we thus consider the Starobinsky potential
\begin{align}
V(\phi) = \frac{3\mu^2M_{\rm p}^2}{4}\left(1-e^{-\sqrt{\frac{2}{3}}\frac{\phi}{M_{\rm p}}}\right)^2
\end{align}
with $\mu=1.14\times10^{-5}M_{\rm p}$ as a typical representative to present the predictions of our model. According to the discussion in Sec. \ref{Model}, we set a nonzero $\chi$ background with $\chi = M_{\rm p}$ at $\phi=5.6M_{\rm p}$ and a heavy mass with $m_{\chi}=0.1\mu$ having the same order of magnitude as $H$. Then, we perform the numerical calculations with the three sets of parameters shown in Table \ref{table1}. 

Taking Set 1 as an example, we first show the evolution of the scalar perturbations with a resonant mode. In Fig. \ref{fig1}, we plot the time evolution of the power spectrum at a resonant scale for each of the perturbations $\delta\chi$, $\delta\phi$, and $\Psi$. This figure indicates that $\delta\phi_k$ exponentially grows as $\delta\chi_k$ exponentially grows, which is the result of the noncanonical kinetic coupling of the $\phi$ and $\chi$ fields. Meanwhile, it is easy to observe that $\delta\chi_k$ decays far outside the horizon due to $m_\chi \sim \mathcal{O}(H)$. Moreover, it is evident that the scalar metric perturbation is a negligible source of GWs relative to the scalar fields.

\begin{figure}
	\centering
	\includegraphics[width=0.8\textwidth ]{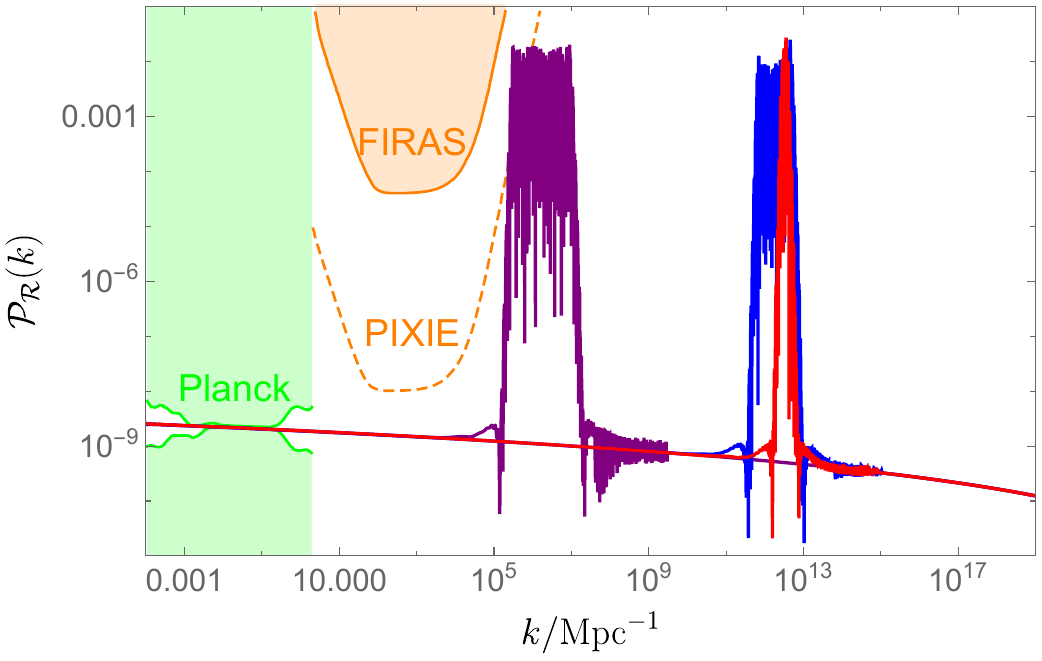}
	\caption{\label{fig2} The power spectra of the curvature perturbations obtained by setting the \textit{e}-folding number from the time when the pivot scale $k_\ast = 0.05\;{\rm Mpc}^{-1}$ exits the horizon to the end of inflation as $60$. The red, blue and purple lines represent the results in Set 1, Set 2, and Set 3, respectively.
	The green shaded region is excluded by the CMB observations \cite{Akrami:2018odb}.  The orange shaded region shows the current upper bound on the power spectrum from measurements of $\mu$ distortion for COBE/FIRAS \cite{Mather:1993ij,Fixsen:1996nj}. The forecasted constraint for the distortion experiment PIXIE \cite{Kogut:2011xw} is shown as the orange dashed line. See Ref. \cite{Chluba:2019nxa} for the summary of constraints on the power spectrum of the curvature perturbations. }
\end{figure}

\begin{figure}
	\centering
	\includegraphics[width=0.48\textwidth ]{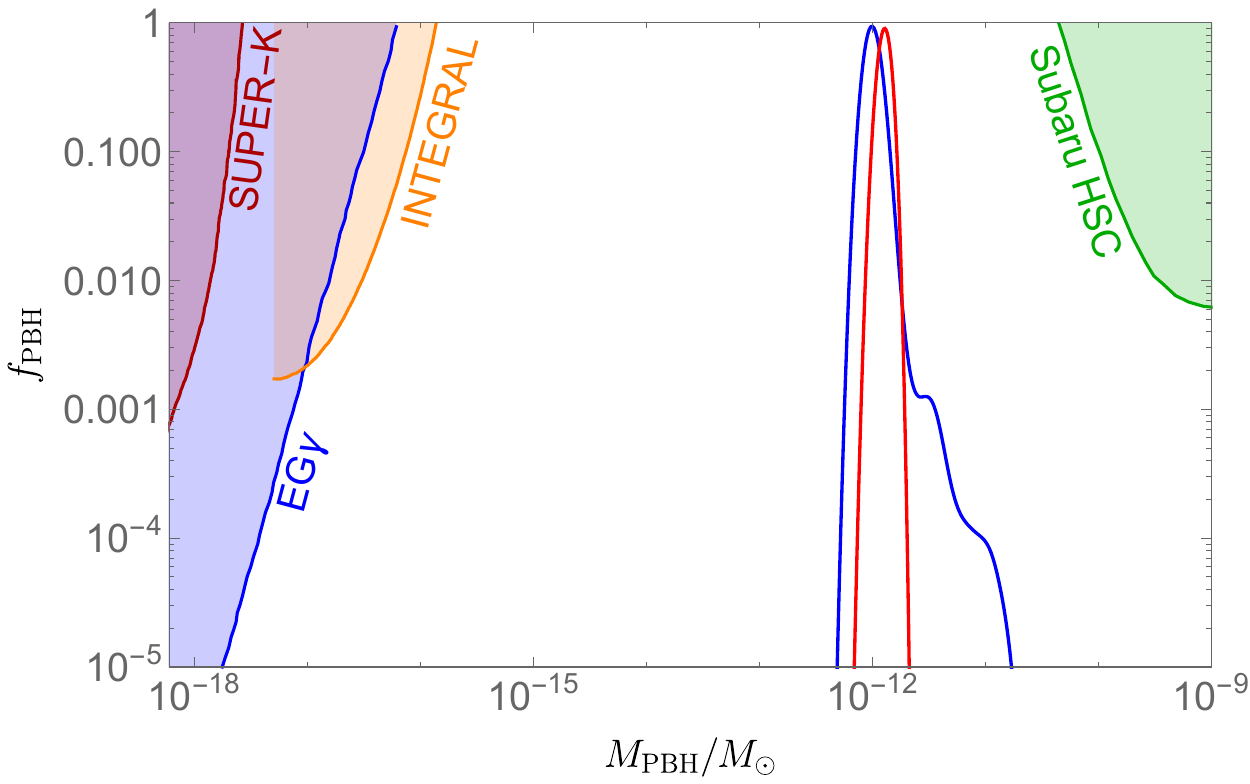}
	\includegraphics[width=0.48\textwidth ]{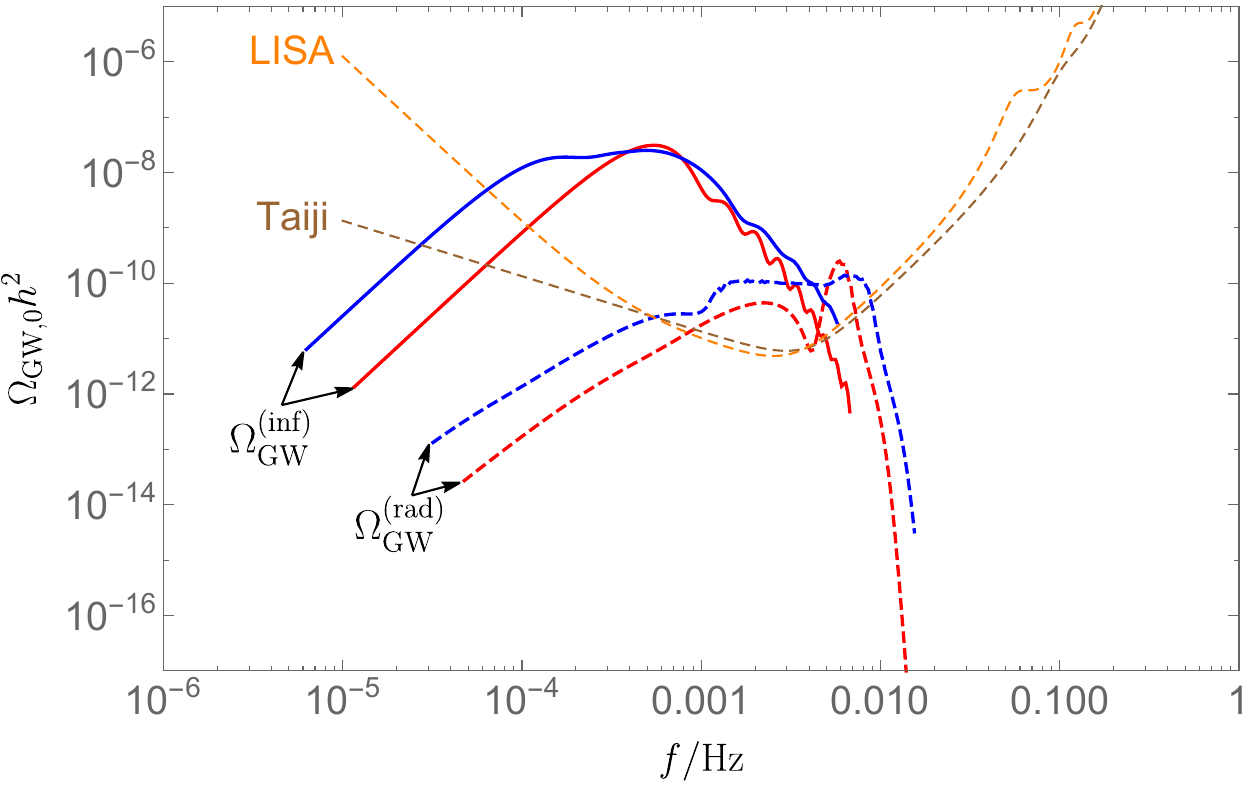}
	\caption{\label{fig3} The mass spectra of PBHs (left panel) and the current energy spectra of the induced GWs (right panel) for Set 1 and Set 2. The red and blue lines correspond to the results of Set 1 and Set 2, respectively. In the left panel, the lines with half shades represent the current observational constrains on PBH abundance: diffuse supernova neutrino background (SUPER-K) \cite{Dasgupta:2019cae}, extra-galactic gamma ray by the Hawking radiation (EG$\gamma$) \cite{Carr:2009jm}, galactic center 511 keV gamma-ray line (INTEGRAL) \cite{Laha:2019ssq}, and Subaru HSC microlensing (Subaru HSC) \cite{Smyth:2019whb}. In the right panel, the orange and brown dashed lines represent the sensitivity curves of the space-based projects LISA \cite{LISA:2017pwj} and Taiji \cite{Ruan:2018tsw}, respectively. }
\end{figure}

Figure \ref{fig2} displays the resulting power spectrum of the curvature perturbations for these three sets of parameters. For Set 1 and Set 2, the resonant scales of the curvature perturbations are both located in the range of around $10^{12}{\rm Mpc}^{-1}$, and the both amplitudes of the corresponding power spectra are of order $10^{-2}$. The main difference between the two sets embodies the width of the amplified part of power spectrum due to the different field excursion $\Delta\phi$. The power spectra of these two cases result in the production of a sizable amount of PBHs with masses around $10^{-12}M_\odot$, as shown in the left panel of Fig. \ref{fig3}. The resulting PBHs in Set 1 and Set 2, comprising $28\%$ and $39\%$ of DM respectively, can make up of the vital component of DM. The predicted total energy spectra of GWs for Set 1 and Set 2 are plotted in the right panel of Fig. \ref{fig3}. It is easy to see that small $\Delta\phi$ (Set 1) and the large $\Delta\phi$ (Set 2) result in a peak and a broad plateau in the energy spectra of GWs from both the inflationary era and the radiation-dominated era, respectively. Interestingly, the height of the peak or plateau in the energy spectrum of GWs from inflation is much larger than that from the radiation-dominated era; however, the frequency of the peak or plateau in the former is located in the lower frequency range relative to that in the latter. As a result, the total energy spectrum of GWs is not dominated by the energy spectrum of GWs from inflation but displays a unique uneven double-peak or double-plateau pattern. In addition, the energy spectra of GWs from both the inflationary era and the radiation-dominated era exceed the sensitivity curves of LISA and Taiji, and the detection of such SGWBs provides us a chance to test our model.

\begin{figure}
	\centering
	\includegraphics[width=0.48\textwidth ]{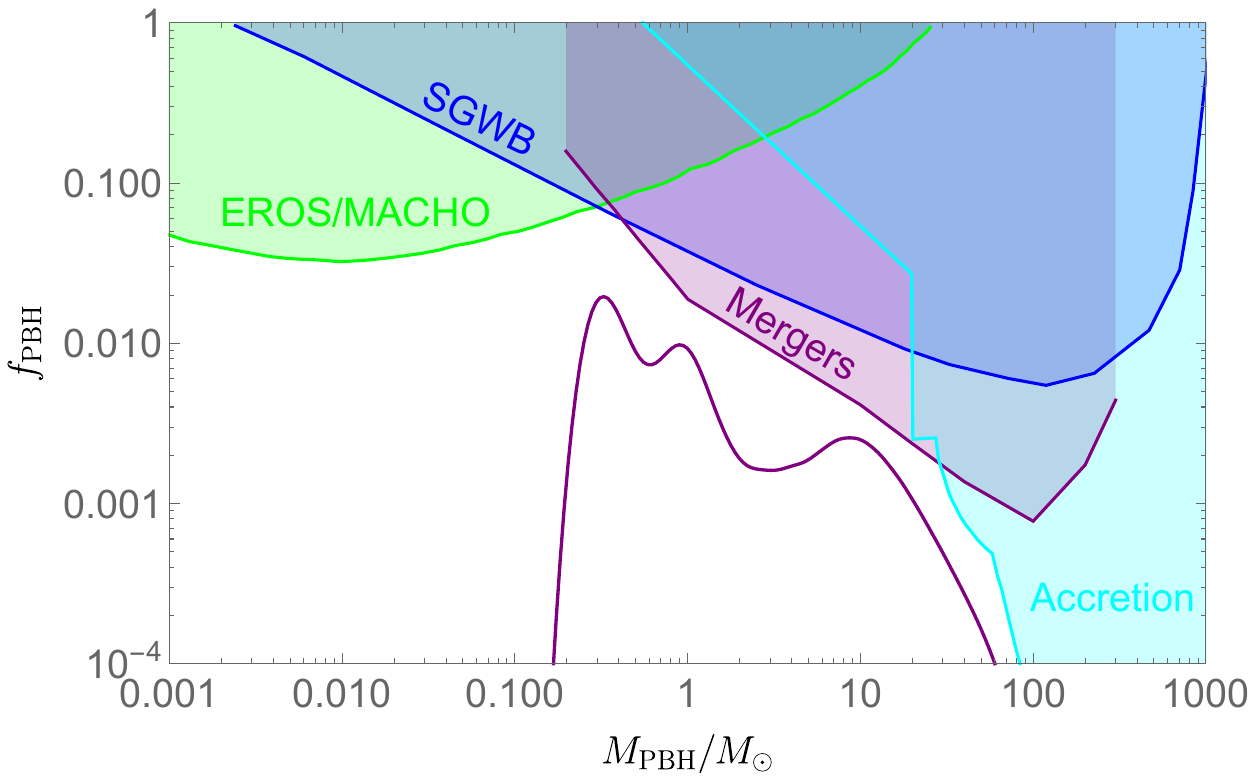}
	\includegraphics[width=0.48\textwidth ]{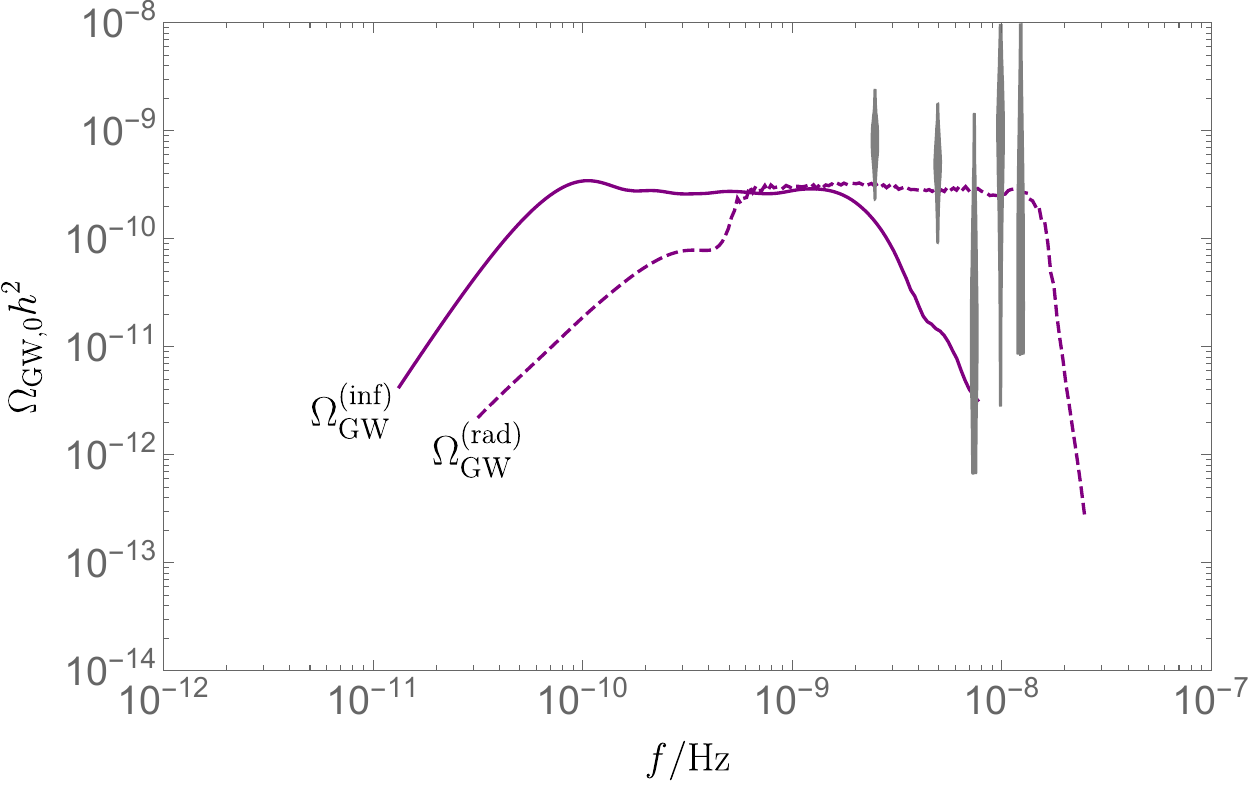}
	\caption{\label{fig4} The mass spectrum of PBHs (left panel) and the current energy spectra of the induced GWs (right panel) for Set 3. In the left panel, the lines with half shades show the  gravitational lensing constraint from EROS/MACHO (EROS/MACHO) \cite{EROS-2:2006ryy}, the constraints from GWs produced by individual mergers (Mergers) \cite{Kavanagh:2018ggo,LIGOScientific:2019kan} and stochastic background of mergers (SGWB) \cite{Chen:2019irf}, and the accretion constraints (Accretion) from CMB \cite{Serpico:2020ehh}, EDGES 21cm \cite{Hektor:2018qqw}, X ray \cite{Manshanden:2018tze}, radio \cite{Manshanden:2018tze}, and dwarf galaxy heating \cite{Lu:2020bmd}. See Refs. \cite{Carr:2020gox,Green:2020jor} for the summary of constraints on the abundance of PBHs. In the right panel, the gray violin plots show the five first frequency bins from the NANOGrav observation \cite{Arzoumanian:2020vkk}. }
\end{figure}

For Set 3, the resonant scales of the power spectrum of the curvature perturbations approximately span from $10^5{\rm Mpc}^{-1}$ to $10^7{\rm Mpc}^{-1}$, as seen in Fig. \ref{fig2}. We can see from the left panel of Fig. \ref{fig4} that $f_{\rm PBH}\sim 10^{-3}$ at $M \sim \mathcal{O}(10)M_\odot$, which may account for the merger rage expected by the LIGO-Virgo Collaboration \cite{Sasaki:2016jop}. We plot the energy spectra of GWs induced during the inflationary era and the radiation-dominated era compared with the five first frequency bins from NANOGrav in the right panel of Fig. \ref{fig4}. Obviously, the induced GWs associated with the PBH formation are compatible with the NANOGrav results. Note that in the choice of $\delta_c \simeq 0.45$, the predicted GWs from the radiation-dominated era are too large to explain the NANOGrav results if we take the Gaussian or $k$-space top-hat window function.

\section{Conclusions}
\label{conclusions}
In this work, we studied the amplification of the curvature perturbations in the single-field slow-roll inflation with a spectator scalar field kinetically coupled to the inflaton. After introducing a periodic coupling function of the inflaton, the perturbations of the spectator field at certain scales are exponentially amplified through the parametric resonance, and then the enhanced spectator field perturbations trigger the amplification of the inflaton perturbations at first-order level due to the kinetic coupling and a nonzero background of the spectator field. As a result, the curvature perturbations are enhanced to lead to the production of a sizable amount of PBHs. Meanwhile, due to ineludible second-order effect of overlarge scalar perturbations, the model leaves a sizable SGWB contributed from the inflationary era and the radiation-dominated era. Choosing the appropriate model parameters, we obtain the mass spectrum of PBHs at $ \mathcal{O}(10^{-12})M_\odot $ as a vital component of DM. The corresponding SGWB, including two components, may be detected by the future space-based projects such as LISA and Taiji, and its energy spectrum displays unique uneven double-peak or double-plateau structure, which can be utilized to test our model. Moreover, for the case of PBHs that could explain the LIGO-Virgo events, we find that the energy spectrum of the resulting GWs originating from the radiation-dominated era is consistent with the NANOGrav signal.

\begin{acknowledgments}
We are grateful to Jing Liu and Lang Liu for useful discussions and comments. This work is supported in part  by the National Key Research and Development Program of China Grant No. 2020YFC2201501, the National Natural Science Foundation of China Grants No. 11947302, No. 11991052, No. 11690022, No. 11821505 and No. 12047559,  the China Postdoctoral Science Foundation Grant No. 2020M680689,  the Key Research Program of the CAS Grant No. XDPB15, and the Key Research Program of Frontier Sciences of CAS.

\end{acknowledgments}


\bibliographystyle{apsrev4-1}
\bibliography{references}

\begin{thebibliography}{119}%
\makeatletter
\providecommand \@ifxundefined [1]{%
 \@ifx{#1\undefined}
}%
\providecommand \@ifnum [1]{%
 \ifnum #1\expandafter \@firstoftwo
 \else \expandafter \@secondoftwo
 \fi
}%
\providecommand \@ifx [1]{%
 \ifx #1\expandafter \@firstoftwo
 \else \expandafter \@secondoftwo
 \fi
}%
\providecommand \natexlab [1]{#1}%
\providecommand \enquote  [1]{``#1''}%
\providecommand \bibnamefont  [1]{#1}%
\providecommand \bibfnamefont [1]{#1}%
\providecommand \citenamefont [1]{#1}%
\providecommand \href@noop [0]{\@secondoftwo}%
\providecommand \href [0]{\begingroup \@sanitize@url \@href}%
\providecommand \@href[1]{\@@startlink{#1}\@@href}%
\providecommand \@@href[1]{\endgroup#1\@@endlink}%
\providecommand \@sanitize@url [0]{\catcode `\\12\catcode `\$12\catcode
  `\&12\catcode `\#12\catcode `\^12\catcode `\_12\catcode `\%12\relax}%
\providecommand \@@startlink[1]{}%
\providecommand \@@endlink[0]{}%
\providecommand \url  [0]{\begingroup\@sanitize@url \@url }%
\providecommand \@url [1]{\endgroup\@href {#1}{\urlprefix }}%
\providecommand \urlprefix  [0]{URL }%
\providecommand \Eprint [0]{\href }%
\providecommand \doibase [0]{http://dx.doi.org/}%
\providecommand \selectlanguage [0]{\@gobble}%
\providecommand \bibinfo  [0]{\@secondoftwo}%
\providecommand \bibfield  [0]{\@secondoftwo}%
\providecommand \translation [1]{[#1]}%
\providecommand \BibitemOpen [0]{}%
\providecommand \bibitemStop [0]{}%
\providecommand \bibitemNoStop [0]{.\EOS\space}%
\providecommand \EOS [0]{\spacefactor3000\relax}%
\providecommand \BibitemShut  [1]{\csname bibitem#1\endcsname}%
\let\auto@bib@innerbib\@empty
\bibitem [{\citenamefont {Hawking}(1971)}]{Hawking:1971ei}%
  \BibitemOpen
  \bibfield  {author} {\bibinfo {author} {\bibfnamefont {S.}~\bibnamefont
  {Hawking}},\ }\href@noop {} {\bibfield  {journal} {\bibinfo  {journal} {Mon.
  Not. Roy. Astron. Soc.}\ }\textbf {\bibinfo {volume} {152}},\ \bibinfo
  {pages} {75} (\bibinfo {year} {1971})}\BibitemShut {NoStop}%
\bibitem [{\citenamefont {Carr}\ and\ \citenamefont
  {Hawking}(1974)}]{Carr:1974nx}%
  \BibitemOpen
  \bibfield  {author} {\bibinfo {author} {\bibfnamefont {B.~J.}\ \bibnamefont
  {Carr}}\ and\ \bibinfo {author} {\bibfnamefont {S.~W.}\ \bibnamefont
  {Hawking}},\ }\href@noop {} {\bibfield  {journal} {\bibinfo  {journal} {Mon.
  Not. Roy. Astron. Soc.}\ }\textbf {\bibinfo {volume} {168}},\ \bibinfo
  {pages} {399} (\bibinfo {year} {1974})}\BibitemShut {NoStop}%
\bibitem [{\citenamefont {Carr}(1975)}]{Carr:1975qj}%
  \BibitemOpen
  \bibfield  {author} {\bibinfo {author} {\bibfnamefont {B.~J.}\ \bibnamefont
  {Carr}},\ }\href {\doibase 10.1086/153853} {\bibfield  {journal} {\bibinfo
  {journal} {Astrophys. J.}\ }\textbf {\bibinfo {volume} {201}},\ \bibinfo
  {pages} {1} (\bibinfo {year} {1975})}\BibitemShut {NoStop}%
\bibitem [{\citenamefont {Belotsky}\ \emph {et~al.}(2014)\citenamefont
  {Belotsky}, \citenamefont {Dmitriev}, \citenamefont {Esipova}, \citenamefont
  {Gani}, \citenamefont {Grobov}, \citenamefont {Khlopov}, \citenamefont
  {Kirillov}, \citenamefont {Rubin},\ and\ \citenamefont
  {Svadkovsky}}]{Belotsky:2014kca}%
  \BibitemOpen
  \bibfield  {author} {\bibinfo {author} {\bibfnamefont {K.~M.}\ \bibnamefont
  {Belotsky}}, \bibinfo {author} {\bibfnamefont {A.~D.}\ \bibnamefont
  {Dmitriev}}, \bibinfo {author} {\bibfnamefont {E.~A.}\ \bibnamefont
  {Esipova}}, \bibinfo {author} {\bibfnamefont {V.~A.}\ \bibnamefont {Gani}},
  \bibinfo {author} {\bibfnamefont {A.~V.}\ \bibnamefont {Grobov}}, \bibinfo
  {author} {\bibfnamefont {M.~Y.}\ \bibnamefont {Khlopov}}, \bibinfo {author}
  {\bibfnamefont {A.~A.}\ \bibnamefont {Kirillov}}, \bibinfo {author}
  {\bibfnamefont {S.~G.}\ \bibnamefont {Rubin}}, \ and\ \bibinfo {author}
  {\bibfnamefont {I.~V.}\ \bibnamefont {Svadkovsky}},\ }\href {\doibase
  10.1142/S0217732314400057} {\bibfield  {journal} {\bibinfo  {journal} {Mod.
  Phys. Lett. A}\ }\textbf {\bibinfo {volume} {29}},\ \bibinfo {pages}
  {1440005} (\bibinfo {year} {2014})},\ \Eprint
  {http://arxiv.org/abs/1410.0203} {arXiv:1410.0203 [astro-ph.CO]} \BibitemShut
  {NoStop}%
\bibitem [{\citenamefont {Carr}\ \emph {et~al.}(2016)\citenamefont {Carr},
  \citenamefont {Kuhnel},\ and\ \citenamefont {Sandstad}}]{Carr:2016drx}%
  \BibitemOpen
  \bibfield  {author} {\bibinfo {author} {\bibfnamefont {B.}~\bibnamefont
  {Carr}}, \bibinfo {author} {\bibfnamefont {F.}~\bibnamefont {Kuhnel}}, \ and\
  \bibinfo {author} {\bibfnamefont {M.}~\bibnamefont {Sandstad}},\ }\href
  {\doibase 10.1103/PhysRevD.94.083504} {\bibfield  {journal} {\bibinfo
  {journal} {Phys. Rev. D}\ }\textbf {\bibinfo {volume} {94}},\ \bibinfo
  {pages} {083504} (\bibinfo {year} {2016})},\ \Eprint
  {http://arxiv.org/abs/1607.06077} {arXiv:1607.06077 [astro-ph.CO]}
  \BibitemShut {NoStop}%
\bibitem [{\citenamefont {Carr}\ and\ \citenamefont
  {Kuhnel}(2020)}]{Carr:2020xqk}%
  \BibitemOpen
  \bibfield  {author} {\bibinfo {author} {\bibfnamefont {B.}~\bibnamefont
  {Carr}}\ and\ \bibinfo {author} {\bibfnamefont {F.}~\bibnamefont {Kuhnel}},\
  }\href {\doibase 10.1146/annurev-nucl-050520-125911} {\bibfield  {journal}
  {\bibinfo  {journal} {Ann. Rev. Nucl. Part. Sci.}\ }\textbf {\bibinfo
  {volume} {70}},\ \bibinfo {pages} {355} (\bibinfo {year} {2020})},\ \Eprint
  {http://arxiv.org/abs/2006.02838} {arXiv:2006.02838 [astro-ph.CO]}
  \BibitemShut {NoStop}%
\bibitem [{\citenamefont {Bird}\ \emph {et~al.}(2016)\citenamefont {Bird},
  \citenamefont {Cholis}, \citenamefont {Mu\~noz}, \citenamefont {Ali-Haimoud},
  \citenamefont {Kamionkowski}, \citenamefont {Kovetz}, \citenamefont
  {Raccanelli},\ and\ \citenamefont {Riess}}]{Bird:2016dcv}%
  \BibitemOpen
  \bibfield  {author} {\bibinfo {author} {\bibfnamefont {S.}~\bibnamefont
  {Bird}}, \bibinfo {author} {\bibfnamefont {I.}~\bibnamefont {Cholis}},
  \bibinfo {author} {\bibfnamefont {J.~B.}\ \bibnamefont {Mu\~noz}}, \bibinfo
  {author} {\bibfnamefont {Y.}~\bibnamefont {Ali-Haimoud}}, \bibinfo {author}
  {\bibfnamefont {M.}~\bibnamefont {Kamionkowski}}, \bibinfo {author}
  {\bibfnamefont {E.~D.}\ \bibnamefont {Kovetz}}, \bibinfo {author}
  {\bibfnamefont {A.}~\bibnamefont {Raccanelli}}, \ and\ \bibinfo {author}
  {\bibfnamefont {A.~G.}\ \bibnamefont {Riess}},\ }\href {\doibase
  10.1103/PhysRevLett.116.201301} {\bibfield  {journal} {\bibinfo  {journal}
  {Phys. Rev. Lett.}\ }\textbf {\bibinfo {volume} {116}},\ \bibinfo {pages}
  {201301} (\bibinfo {year} {2016})},\ \Eprint
  {http://arxiv.org/abs/1603.00464} {arXiv:1603.00464 [astro-ph.CO]}
  \BibitemShut {NoStop}%
\bibitem [{\citenamefont {Sasaki}\ \emph {et~al.}(2016)\citenamefont {Sasaki},
  \citenamefont {Suyama}, \citenamefont {Tanaka},\ and\ \citenamefont
  {Yokoyama}}]{Sasaki:2016jop}%
  \BibitemOpen
  \bibfield  {author} {\bibinfo {author} {\bibfnamefont {M.}~\bibnamefont
  {Sasaki}}, \bibinfo {author} {\bibfnamefont {T.}~\bibnamefont {Suyama}},
  \bibinfo {author} {\bibfnamefont {T.}~\bibnamefont {Tanaka}}, \ and\ \bibinfo
  {author} {\bibfnamefont {S.}~\bibnamefont {Yokoyama}},\ }\href {\doibase
  10.1103/PhysRevLett.117.061101} {\bibfield  {journal} {\bibinfo  {journal}
  {Phys. Rev. Lett.}\ }\textbf {\bibinfo {volume} {117}},\ \bibinfo {pages}
  {061101} (\bibinfo {year} {2016})},\ \bibinfo {note} {[Erratum:
  Phys.Rev.Lett. 121, 059901(E) (2018)]},\ \Eprint
  {http://arxiv.org/abs/1603.08338} {arXiv:1603.08338 [astro-ph.CO]}
  \BibitemShut {NoStop}%
\bibitem [{\citenamefont {Abbott}\ \emph
  {et~al.}(2016{\natexlab{a}})\citenamefont {Abbott} \emph
  {et~al.}}]{Abbott:2016blz}%
  \BibitemOpen
  \bibfield  {author} {\bibinfo {author} {\bibfnamefont {B.~P.}\ \bibnamefont
  {Abbott}} \emph {et~al.} (\bibinfo {collaboration} {LIGO Scientific,
  Virgo}),\ }\href {\doibase 10.1103/PhysRevLett.116.061102} {\bibfield
  {journal} {\bibinfo  {journal} {Phys. Rev. Lett.}\ }\textbf {\bibinfo
  {volume} {116}},\ \bibinfo {pages} {061102} (\bibinfo {year}
  {2016}{\natexlab{a}})},\ \Eprint {http://arxiv.org/abs/1602.03837}
  {arXiv:1602.03837 [gr-qc]} \BibitemShut {NoStop}%
\bibitem [{\citenamefont {Abbott}\ \emph
  {et~al.}(2016{\natexlab{b}})\citenamefont {Abbott} \emph
  {et~al.}}]{Abbott:2016nmj}%
  \BibitemOpen
  \bibfield  {author} {\bibinfo {author} {\bibfnamefont {B.~P.}\ \bibnamefont
  {Abbott}} \emph {et~al.} (\bibinfo {collaboration} {LIGO Scientific,
  Virgo}),\ }\href {\doibase 10.1103/PhysRevLett.116.241103} {\bibfield
  {journal} {\bibinfo  {journal} {Phys. Rev. Lett.}\ }\textbf {\bibinfo
  {volume} {116}},\ \bibinfo {pages} {241103} (\bibinfo {year}
  {2016}{\natexlab{b}})},\ \Eprint {http://arxiv.org/abs/1606.04855}
  {arXiv:1606.04855 [gr-qc]} \BibitemShut {NoStop}%
\bibitem [{\citenamefont {Abbott}\ \emph {et~al.}(2017)\citenamefont {Abbott}
  \emph {et~al.}}]{Abbott:2017vtc}%
  \BibitemOpen
  \bibfield  {author} {\bibinfo {author} {\bibfnamefont {B.~P.}\ \bibnamefont
  {Abbott}} \emph {et~al.} (\bibinfo {collaboration} {LIGO Scientific,
  VIRGO}),\ }\href {\doibase 10.1103/PhysRevLett.118.221101} {\bibfield
  {journal} {\bibinfo  {journal} {Phys. Rev. Lett.}\ }\textbf {\bibinfo
  {volume} {118}},\ \bibinfo {pages} {221101} (\bibinfo {year} {2017})},\
  \bibinfo {note} {[Erratum: Phys.Rev.Lett. 121, 129901 (2018)]},\ \Eprint
  {http://arxiv.org/abs/1706.01812} {arXiv:1706.01812 [gr-qc]} \BibitemShut
  {NoStop}%
\bibitem [{\citenamefont {Mukherjee}\ and\ \citenamefont
  {Silk}(2021)}]{Mukherjee:2021ags}%
  \BibitemOpen
  \bibfield  {author} {\bibinfo {author} {\bibfnamefont {S.}~\bibnamefont
  {Mukherjee}}\ and\ \bibinfo {author} {\bibfnamefont {J.}~\bibnamefont
  {Silk}},\ }\href {\doibase 10.1093/mnras/stab1932} {\bibfield  {journal}
  {\bibinfo  {journal} {Mon. Not. Roy. Astron. Soc.}\ }\textbf {\bibinfo
  {volume} {506}},\ \bibinfo {pages} {3977} (\bibinfo {year} {2021})},\ \Eprint
  {http://arxiv.org/abs/2105.11139} {arXiv:2105.11139 [gr-qc]} \BibitemShut
  {NoStop}%
\bibitem [{\citenamefont {Mukherjee}\ \emph {et~al.}(2021)\citenamefont
  {Mukherjee}, \citenamefont {Meinema},\ and\ \citenamefont
  {Silk}}]{Mukherjee:2021itf}%
  \BibitemOpen
  \bibfield  {author} {\bibinfo {author} {\bibfnamefont {S.}~\bibnamefont
  {Mukherjee}}, \bibinfo {author} {\bibfnamefont {M.~S.~P.}\ \bibnamefont
  {Meinema}}, \ and\ \bibinfo {author} {\bibfnamefont {J.}~\bibnamefont
  {Silk}},\ }\href@noop {} {\  (\bibinfo {year} {2021})},\ \Eprint
  {http://arxiv.org/abs/2107.02181} {arXiv:2107.02181 [astro-ph.CO]}
  \BibitemShut {NoStop}%
\bibitem [{\citenamefont {Matarrese}\ \emph {et~al.}(1998)\citenamefont
  {Matarrese}, \citenamefont {Mollerach},\ and\ \citenamefont
  {Bruni}}]{Matarrese:1997ay}%
  \BibitemOpen
  \bibfield  {author} {\bibinfo {author} {\bibfnamefont {S.}~\bibnamefont
  {Matarrese}}, \bibinfo {author} {\bibfnamefont {S.}~\bibnamefont
  {Mollerach}}, \ and\ \bibinfo {author} {\bibfnamefont {M.}~\bibnamefont
  {Bruni}},\ }\href {\doibase 10.1103/PhysRevD.58.043504} {\bibfield  {journal}
  {\bibinfo  {journal} {Phys. Rev. D}\ }\textbf {\bibinfo {volume} {58}},\
  \bibinfo {pages} {043504} (\bibinfo {year} {1998})},\ \Eprint
  {http://arxiv.org/abs/astro-ph/9707278} {arXiv:astro-ph/9707278} \BibitemShut
  {NoStop}%
\bibitem [{\citenamefont {Noh}\ and\ \citenamefont {Hwang}(2004)}]{Noh:2004bc}%
  \BibitemOpen
  \bibfield  {author} {\bibinfo {author} {\bibfnamefont {H.}~\bibnamefont
  {Noh}}\ and\ \bibinfo {author} {\bibfnamefont {J.-c.}\ \bibnamefont
  {Hwang}},\ }\href {\doibase 10.1103/PhysRevD.69.104011} {\bibfield  {journal}
  {\bibinfo  {journal} {Phys. Rev. D}\ }\textbf {\bibinfo {volume} {69}},\
  \bibinfo {pages} {104011} (\bibinfo {year} {2004})}\BibitemShut {NoStop}%
\bibitem [{\citenamefont {Ananda}\ \emph {et~al.}(2007)\citenamefont {Ananda},
  \citenamefont {Clarkson},\ and\ \citenamefont {Wands}}]{Ananda:2006af}%
  \BibitemOpen
  \bibfield  {author} {\bibinfo {author} {\bibfnamefont {K.~N.}\ \bibnamefont
  {Ananda}}, \bibinfo {author} {\bibfnamefont {C.}~\bibnamefont {Clarkson}}, \
  and\ \bibinfo {author} {\bibfnamefont {D.}~\bibnamefont {Wands}},\ }\href
  {\doibase 10.1103/PhysRevD.75.123518} {\bibfield  {journal} {\bibinfo
  {journal} {Phys. Rev. D}\ }\textbf {\bibinfo {volume} {75}},\ \bibinfo
  {pages} {123518} (\bibinfo {year} {2007})},\ \Eprint
  {http://arxiv.org/abs/gr-qc/0612013} {arXiv:gr-qc/0612013} \BibitemShut
  {NoStop}%
\bibitem [{\citenamefont {Baumann}\ \emph {et~al.}(2007)\citenamefont
  {Baumann}, \citenamefont {Steinhardt}, \citenamefont {Takahashi},\ and\
  \citenamefont {Ichiki}}]{Baumann:2007zm}%
  \BibitemOpen
  \bibfield  {author} {\bibinfo {author} {\bibfnamefont {D.}~\bibnamefont
  {Baumann}}, \bibinfo {author} {\bibfnamefont {P.~J.}\ \bibnamefont
  {Steinhardt}}, \bibinfo {author} {\bibfnamefont {K.}~\bibnamefont
  {Takahashi}}, \ and\ \bibinfo {author} {\bibfnamefont {K.}~\bibnamefont
  {Ichiki}},\ }\href {\doibase 10.1103/PhysRevD.76.084019} {\bibfield
  {journal} {\bibinfo  {journal} {Phys. Rev. D}\ }\textbf {\bibinfo {volume}
  {76}},\ \bibinfo {pages} {084019} (\bibinfo {year} {2007})},\ \Eprint
  {http://arxiv.org/abs/hep-th/0703290} {arXiv:hep-th/0703290} \BibitemShut
  {NoStop}%
\bibitem [{\citenamefont {Wang}\ and\ \citenamefont
  {Zhang}(2017)}]{Wang:2017krj}%
  \BibitemOpen
  \bibfield  {author} {\bibinfo {author} {\bibfnamefont {B.}~\bibnamefont
  {Wang}}\ and\ \bibinfo {author} {\bibfnamefont {Y.}~\bibnamefont {Zhang}},\
  }\href {\doibase 10.1103/PhysRevD.96.103522} {\bibfield  {journal} {\bibinfo
  {journal} {Phys. Rev. D}\ }\textbf {\bibinfo {volume} {96}},\ \bibinfo
  {pages} {103522} (\bibinfo {year} {2017})},\ \Eprint
  {http://arxiv.org/abs/1710.06641} {arXiv:1710.06641 [gr-qc]} \BibitemShut
  {NoStop}%
\bibitem [{\citenamefont {Dom\`enech}\ and\ \citenamefont
  {Sasaki}(2018)}]{Domenech:2017ems}%
  \BibitemOpen
  \bibfield  {author} {\bibinfo {author} {\bibfnamefont {G.}~\bibnamefont
  {Dom\`enech}}\ and\ \bibinfo {author} {\bibfnamefont {M.}~\bibnamefont
  {Sasaki}},\ }\href {\doibase 10.1103/PhysRevD.97.023521} {\bibfield
  {journal} {\bibinfo  {journal} {Phys. Rev. D}\ }\textbf {\bibinfo {volume}
  {97}},\ \bibinfo {pages} {023521} (\bibinfo {year} {2018})},\ \Eprint
  {http://arxiv.org/abs/1709.09804} {arXiv:1709.09804 [gr-qc]} \BibitemShut
  {NoStop}%
\bibitem [{\citenamefont {Saito}\ and\ \citenamefont
  {Yokoyama}(2009)}]{Saito:2008jc}%
  \BibitemOpen
  \bibfield  {author} {\bibinfo {author} {\bibfnamefont {R.}~\bibnamefont
  {Saito}}\ and\ \bibinfo {author} {\bibfnamefont {J.}~\bibnamefont
  {Yokoyama}},\ }\href {\doibase 10.1103/PhysRevLett.102.161101} {\bibfield
  {journal} {\bibinfo  {journal} {Phys. Rev. Lett.}\ }\textbf {\bibinfo
  {volume} {102}},\ \bibinfo {pages} {161101} (\bibinfo {year} {2009})},\
  \bibinfo {note} {[Erratum: Phys.Rev.Lett. 107, 069901(E) (2011)]},\ \Eprint
  {http://arxiv.org/abs/0812.4339} {arXiv:0812.4339 [astro-ph]} \BibitemShut
  {NoStop}%
\bibitem [{\citenamefont {Saito}\ and\ \citenamefont
  {Yokoyama}(2010)}]{Saito:2009jt}%
  \BibitemOpen
  \bibfield  {author} {\bibinfo {author} {\bibfnamefont {R.}~\bibnamefont
  {Saito}}\ and\ \bibinfo {author} {\bibfnamefont {J.}~\bibnamefont
  {Yokoyama}},\ }\href {\doibase 10.1143/PTP.126.351} {\bibfield  {journal}
  {\bibinfo  {journal} {Prog. Theor. Phys.}\ }\textbf {\bibinfo {volume}
  {123}},\ \bibinfo {pages} {867} (\bibinfo {year} {2010})},\ \bibinfo {note}
  {[Erratum: Prog.Theor.Phys. 126, 351--352 (2011)]},\ \Eprint
  {http://arxiv.org/abs/0912.5317} {arXiv:0912.5317 [astro-ph.CO]} \BibitemShut
  {NoStop}%
\bibitem [{\citenamefont {Bugaev}\ and\ \citenamefont
  {Klimai}(2011)}]{Bugaev:2010bb}%
  \BibitemOpen
  \bibfield  {author} {\bibinfo {author} {\bibfnamefont {E.}~\bibnamefont
  {Bugaev}}\ and\ \bibinfo {author} {\bibfnamefont {P.}~\bibnamefont
  {Klimai}},\ }\href {\doibase 10.1103/PhysRevD.83.083521} {\bibfield
  {journal} {\bibinfo  {journal} {Phys. Rev. D}\ }\textbf {\bibinfo {volume}
  {83}},\ \bibinfo {pages} {083521} (\bibinfo {year} {2011})},\ \Eprint
  {http://arxiv.org/abs/1012.4697} {arXiv:1012.4697 [astro-ph.CO]} \BibitemShut
  {NoStop}%
\bibitem [{\citenamefont {Garcia-Bellido}\ \emph {et~al.}(2017)\citenamefont
  {Garcia-Bellido}, \citenamefont {Peloso},\ and\ \citenamefont
  {Unal}}]{Garcia-Bellido:2017aan}%
  \BibitemOpen
  \bibfield  {author} {\bibinfo {author} {\bibfnamefont {J.}~\bibnamefont
  {Garcia-Bellido}}, \bibinfo {author} {\bibfnamefont {M.}~\bibnamefont
  {Peloso}}, \ and\ \bibinfo {author} {\bibfnamefont {C.}~\bibnamefont
  {Unal}},\ }\href {\doibase 10.1088/1475-7516/2017/09/013} {\bibfield
  {journal} {\bibinfo  {journal} {JCAP}\ }\textbf {\bibinfo {volume} {09}},\
  \bibinfo {pages} {013} (\bibinfo {year} {2017})},\ \Eprint
  {http://arxiv.org/abs/1707.02441} {arXiv:1707.02441 [astro-ph.CO]}
  \BibitemShut {NoStop}%
\bibitem [{\citenamefont {Espinosa}\ \emph {et~al.}(2018)\citenamefont
  {Espinosa}, \citenamefont {Racco},\ and\ \citenamefont
  {Riotto}}]{Espinosa:2018eve}%
  \BibitemOpen
  \bibfield  {author} {\bibinfo {author} {\bibfnamefont {J.~R.}\ \bibnamefont
  {Espinosa}}, \bibinfo {author} {\bibfnamefont {D.}~\bibnamefont {Racco}}, \
  and\ \bibinfo {author} {\bibfnamefont {A.}~\bibnamefont {Riotto}},\ }\href
  {\doibase 10.1088/1475-7516/2018/09/012} {\bibfield  {journal} {\bibinfo
  {journal} {JCAP}\ }\textbf {\bibinfo {volume} {09}},\ \bibinfo {pages} {012}
  (\bibinfo {year} {2018})},\ \Eprint {http://arxiv.org/abs/1804.07732}
  {arXiv:1804.07732 [hep-ph]} \BibitemShut {NoStop}%
\bibitem [{\citenamefont {Kohri}\ and\ \citenamefont
  {Terada}(2018)}]{Kohri:2018awv}%
  \BibitemOpen
  \bibfield  {author} {\bibinfo {author} {\bibfnamefont {K.}~\bibnamefont
  {Kohri}}\ and\ \bibinfo {author} {\bibfnamefont {T.}~\bibnamefont {Terada}},\
  }\href {\doibase 10.1103/PhysRevD.97.123532} {\bibfield  {journal} {\bibinfo
  {journal} {Phys. Rev. D}\ }\textbf {\bibinfo {volume} {97}},\ \bibinfo
  {pages} {123532} (\bibinfo {year} {2018})},\ \Eprint
  {http://arxiv.org/abs/1804.08577} {arXiv:1804.08577 [gr-qc]} \BibitemShut
  {NoStop}%
\bibitem [{\citenamefont {Bartolo}\ \emph {et~al.}(2019)\citenamefont
  {Bartolo}, \citenamefont {De~Luca}, \citenamefont {Franciolini},
  \citenamefont {Peloso}, \citenamefont {Racco},\ and\ \citenamefont
  {Riotto}}]{Bartolo:2018rku}%
  \BibitemOpen
  \bibfield  {author} {\bibinfo {author} {\bibfnamefont {N.}~\bibnamefont
  {Bartolo}}, \bibinfo {author} {\bibfnamefont {V.}~\bibnamefont {De~Luca}},
  \bibinfo {author} {\bibfnamefont {G.}~\bibnamefont {Franciolini}}, \bibinfo
  {author} {\bibfnamefont {M.}~\bibnamefont {Peloso}}, \bibinfo {author}
  {\bibfnamefont {D.}~\bibnamefont {Racco}}, \ and\ \bibinfo {author}
  {\bibfnamefont {A.}~\bibnamefont {Riotto}},\ }\href {\doibase
  10.1103/PhysRevD.99.103521} {\bibfield  {journal} {\bibinfo  {journal} {Phys.
  Rev. D}\ }\textbf {\bibinfo {volume} {99}},\ \bibinfo {pages} {103521}
  (\bibinfo {year} {2019})},\ \Eprint {http://arxiv.org/abs/1810.12224}
  {arXiv:1810.12224 [astro-ph.CO]} \BibitemShut {NoStop}%
\bibitem [{\citenamefont {Cai}\ \emph {et~al.}(2019{\natexlab{a}})\citenamefont
  {Cai}, \citenamefont {Pi},\ and\ \citenamefont {Sasaki}}]{Cai:2018dig}%
  \BibitemOpen
  \bibfield  {author} {\bibinfo {author} {\bibfnamefont {R.-G.}\ \bibnamefont
  {Cai}}, \bibinfo {author} {\bibfnamefont {S.}~\bibnamefont {Pi}}, \ and\
  \bibinfo {author} {\bibfnamefont {M.}~\bibnamefont {Sasaki}},\ }\href
  {\doibase 10.1103/PhysRevLett.122.201101} {\bibfield  {journal} {\bibinfo
  {journal} {Phys. Rev. Lett.}\ }\textbf {\bibinfo {volume} {122}},\ \bibinfo
  {pages} {201101} (\bibinfo {year} {2019}{\natexlab{a}})},\ \Eprint
  {http://arxiv.org/abs/1810.11000} {arXiv:1810.11000 [astro-ph.CO]}
  \BibitemShut {NoStop}%
\bibitem [{\citenamefont {Cai}\ \emph {et~al.}(2019{\natexlab{b}})\citenamefont
  {Cai}, \citenamefont {Chen}, \citenamefont {Tong}, \citenamefont {Wang},\
  and\ \citenamefont {Yan}}]{Cai:2019jah}%
  \BibitemOpen
  \bibfield  {author} {\bibinfo {author} {\bibfnamefont {Y.-F.}\ \bibnamefont
  {Cai}}, \bibinfo {author} {\bibfnamefont {C.}~\bibnamefont {Chen}}, \bibinfo
  {author} {\bibfnamefont {X.}~\bibnamefont {Tong}}, \bibinfo {author}
  {\bibfnamefont {D.-G.}\ \bibnamefont {Wang}}, \ and\ \bibinfo {author}
  {\bibfnamefont {S.-F.}\ \bibnamefont {Yan}},\ }\href {\doibase
  10.1103/PhysRevD.100.043518} {\bibfield  {journal} {\bibinfo  {journal}
  {Phys. Rev. D}\ }\textbf {\bibinfo {volume} {100}},\ \bibinfo {pages}
  {043518} (\bibinfo {year} {2019}{\natexlab{b}})},\ \Eprint
  {http://arxiv.org/abs/1902.08187} {arXiv:1902.08187 [astro-ph.CO]}
  \BibitemShut {NoStop}%
\bibitem [{\citenamefont {Fu}\ \emph {et~al.}(2020{\natexlab{a}})\citenamefont
  {Fu}, \citenamefont {Wu},\ and\ \citenamefont {Yu}}]{Fu:2019vqc}%
  \BibitemOpen
  \bibfield  {author} {\bibinfo {author} {\bibfnamefont {C.}~\bibnamefont
  {Fu}}, \bibinfo {author} {\bibfnamefont {P.}~\bibnamefont {Wu}}, \ and\
  \bibinfo {author} {\bibfnamefont {H.}~\bibnamefont {Yu}},\ }\href {\doibase
  10.1103/PhysRevD.101.023529} {\bibfield  {journal} {\bibinfo  {journal}
  {Phys. Rev. D}\ }\textbf {\bibinfo {volume} {101}},\ \bibinfo {pages}
  {023529} (\bibinfo {year} {2020}{\natexlab{a}})},\ \Eprint
  {http://arxiv.org/abs/1912.05927} {arXiv:1912.05927 [astro-ph.CO]}
  \BibitemShut {NoStop}%
\bibitem [{\citenamefont {Cai}\ \emph {et~al.}(2019{\natexlab{c}})\citenamefont
  {Cai}, \citenamefont {Pi}, \citenamefont {Wang},\ and\ \citenamefont
  {Yang}}]{Cai:2019amo}%
  \BibitemOpen
  \bibfield  {author} {\bibinfo {author} {\bibfnamefont {R.-G.}\ \bibnamefont
  {Cai}}, \bibinfo {author} {\bibfnamefont {S.}~\bibnamefont {Pi}}, \bibinfo
  {author} {\bibfnamefont {S.-J.}\ \bibnamefont {Wang}}, \ and\ \bibinfo
  {author} {\bibfnamefont {X.-Y.}\ \bibnamefont {Yang}},\ }\href {\doibase
  10.1088/1475-7516/2019/05/013} {\bibfield  {journal} {\bibinfo  {journal}
  {JCAP}\ }\textbf {\bibinfo {volume} {05}},\ \bibinfo {pages} {013} (\bibinfo
  {year} {2019}{\natexlab{c}})},\ \Eprint {http://arxiv.org/abs/1901.10152}
  {arXiv:1901.10152 [astro-ph.CO]} \BibitemShut {NoStop}%
\bibitem [{\citenamefont {Pi}\ \emph {et~al.}(2019)\citenamefont {Pi},
  \citenamefont {Sasaki},\ and\ \citenamefont {Zhang}}]{Pi:2019ihn}%
  \BibitemOpen
  \bibfield  {author} {\bibinfo {author} {\bibfnamefont {S.}~\bibnamefont
  {Pi}}, \bibinfo {author} {\bibfnamefont {M.}~\bibnamefont {Sasaki}}, \ and\
  \bibinfo {author} {\bibfnamefont {Y.-l.}\ \bibnamefont {Zhang}},\ }\href
  {\doibase 10.1088/1475-7516/2019/06/049} {\bibfield  {journal} {\bibinfo
  {journal} {JCAP}\ }\textbf {\bibinfo {volume} {06}},\ \bibinfo {pages} {049}
  (\bibinfo {year} {2019})},\ \Eprint {http://arxiv.org/abs/1904.06304}
  {arXiv:1904.06304 [gr-qc]} \BibitemShut {NoStop}%
\bibitem [{\citenamefont {Cai}\ \emph {et~al.}(2019{\natexlab{d}})\citenamefont
  {Cai}, \citenamefont {Pi}, \citenamefont {Wang},\ and\ \citenamefont
  {Yang}}]{Cai:2019elf}%
  \BibitemOpen
  \bibfield  {author} {\bibinfo {author} {\bibfnamefont {R.-G.}\ \bibnamefont
  {Cai}}, \bibinfo {author} {\bibfnamefont {S.}~\bibnamefont {Pi}}, \bibinfo
  {author} {\bibfnamefont {S.-J.}\ \bibnamefont {Wang}}, \ and\ \bibinfo
  {author} {\bibfnamefont {X.-Y.}\ \bibnamefont {Yang}},\ }\href {\doibase
  10.1088/1475-7516/2019/10/059} {\bibfield  {journal} {\bibinfo  {journal}
  {JCAP}\ }\textbf {\bibinfo {volume} {10}},\ \bibinfo {pages} {059} (\bibinfo
  {year} {2019}{\natexlab{d}})},\ \Eprint {http://arxiv.org/abs/1907.06372}
  {arXiv:1907.06372 [astro-ph.CO]} \BibitemShut {NoStop}%
\bibitem [{\citenamefont {Inomata}(2021)}]{Inomata:2020cck}%
  \BibitemOpen
  \bibfield  {author} {\bibinfo {author} {\bibfnamefont {K.}~\bibnamefont
  {Inomata}},\ }\href {\doibase 10.1088/1475-7516/2021/03/013} {\bibfield
  {journal} {\bibinfo  {journal} {JCAP}\ }\textbf {\bibinfo {volume} {03}},\
  \bibinfo {pages} {013} (\bibinfo {year} {2021})},\ \Eprint
  {http://arxiv.org/abs/2008.12300} {arXiv:2008.12300 [gr-qc]} \BibitemShut
  {NoStop}%
\bibitem [{\citenamefont {Pi}\ and\ \citenamefont {Sasaki}(2020)}]{Pi:2020otn}%
  \BibitemOpen
  \bibfield  {author} {\bibinfo {author} {\bibfnamefont {S.}~\bibnamefont
  {Pi}}\ and\ \bibinfo {author} {\bibfnamefont {M.}~\bibnamefont {Sasaki}},\
  }\href {\doibase 10.1088/1475-7516/2020/09/037} {\bibfield  {journal}
  {\bibinfo  {journal} {JCAP}\ }\textbf {\bibinfo {volume} {09}},\ \bibinfo
  {pages} {037} (\bibinfo {year} {2020})},\ \Eprint
  {http://arxiv.org/abs/2005.12306} {arXiv:2005.12306 [gr-qc]} \BibitemShut
  {NoStop}%
\bibitem [{\citenamefont {Braglia}\ \emph {et~al.}(2020)\citenamefont
  {Braglia}, \citenamefont {Hazra}, \citenamefont {Finelli}, \citenamefont
  {Smoot}, \citenamefont {Sriramkumar},\ and\ \citenamefont
  {Starobinsky}}]{Braglia:2020eai}%
  \BibitemOpen
  \bibfield  {author} {\bibinfo {author} {\bibfnamefont {M.}~\bibnamefont
  {Braglia}}, \bibinfo {author} {\bibfnamefont {D.~K.}\ \bibnamefont {Hazra}},
  \bibinfo {author} {\bibfnamefont {F.}~\bibnamefont {Finelli}}, \bibinfo
  {author} {\bibfnamefont {G.~F.}\ \bibnamefont {Smoot}}, \bibinfo {author}
  {\bibfnamefont {L.}~\bibnamefont {Sriramkumar}}, \ and\ \bibinfo {author}
  {\bibfnamefont {A.~A.}\ \bibnamefont {Starobinsky}},\ }\href {\doibase
  10.1088/1475-7516/2020/08/001} {\bibfield  {journal} {\bibinfo  {journal}
  {JCAP}\ }\textbf {\bibinfo {volume} {08}},\ \bibinfo {pages} {001} (\bibinfo
  {year} {2020})},\ \Eprint {http://arxiv.org/abs/2005.02895} {arXiv:2005.02895
  [astro-ph.CO]} \BibitemShut {NoStop}%
\bibitem [{\citenamefont {Braglia}\ \emph {et~al.}(2021)\citenamefont
  {Braglia}, \citenamefont {Chen},\ and\ \citenamefont
  {Hazra}}]{Braglia:2020taf}%
  \BibitemOpen
  \bibfield  {author} {\bibinfo {author} {\bibfnamefont {M.}~\bibnamefont
  {Braglia}}, \bibinfo {author} {\bibfnamefont {X.}~\bibnamefont {Chen}}, \
  and\ \bibinfo {author} {\bibfnamefont {D.~K.}\ \bibnamefont {Hazra}},\ }\href
  {\doibase 10.1088/1475-7516/2021/03/005} {\bibfield  {journal} {\bibinfo
  {journal} {JCAP}\ }\textbf {\bibinfo {volume} {03}},\ \bibinfo {pages} {005}
  (\bibinfo {year} {2021})},\ \Eprint {http://arxiv.org/abs/2012.05821}
  {arXiv:2012.05821 [astro-ph.CO]} \BibitemShut {NoStop}%
\bibitem [{\citenamefont {Peng}\ \emph {et~al.}(2021)\citenamefont {Peng},
  \citenamefont {Fu}, \citenamefont {Liu}, \citenamefont {Guo},\ and\
  \citenamefont {Cai}}]{Peng:2021zon}%
  \BibitemOpen
  \bibfield  {author} {\bibinfo {author} {\bibfnamefont {Z.-Z.}\ \bibnamefont
  {Peng}}, \bibinfo {author} {\bibfnamefont {C.}~\bibnamefont {Fu}}, \bibinfo
  {author} {\bibfnamefont {J.}~\bibnamefont {Liu}}, \bibinfo {author}
  {\bibfnamefont {Z.-K.}\ \bibnamefont {Guo}}, \ and\ \bibinfo {author}
  {\bibfnamefont {R.-G.}\ \bibnamefont {Cai}},\ }\href@noop {} {\  (\bibinfo
  {year} {2021})},\ \Eprint {http://arxiv.org/abs/2106.11816} {arXiv:2106.11816
  [astro-ph.CO]} \BibitemShut {NoStop}%
\bibitem [{\citenamefont {Amaro-Seoane}\ \emph {et~al.}(2017)\citenamefont
  {Amaro-Seoane} \emph {et~al.}}]{LISA:2017pwj}%
  \BibitemOpen
  \bibfield  {author} {\bibinfo {author} {\bibfnamefont {P.}~\bibnamefont
  {Amaro-Seoane}} \emph {et~al.} (\bibinfo {collaboration} {LISA}),\
  }\href@noop {} {\  (\bibinfo {year} {2017})},\ \Eprint
  {http://arxiv.org/abs/1702.00786} {arXiv:1702.00786 [astro-ph.IM]}
  \BibitemShut {NoStop}%
\bibitem [{\citenamefont {Ruan}\ \emph {et~al.}(2020)\citenamefont {Ruan},
  \citenamefont {Guo}, \citenamefont {Cai},\ and\ \citenamefont
  {Zhang}}]{Ruan:2018tsw}%
  \BibitemOpen
  \bibfield  {author} {\bibinfo {author} {\bibfnamefont {W.-H.}\ \bibnamefont
  {Ruan}}, \bibinfo {author} {\bibfnamefont {Z.-K.}\ \bibnamefont {Guo}},
  \bibinfo {author} {\bibfnamefont {R.-G.}\ \bibnamefont {Cai}}, \ and\
  \bibinfo {author} {\bibfnamefont {Y.-Z.}\ \bibnamefont {Zhang}},\ }\href
  {\doibase 10.1142/S0217751X2050075X} {\bibfield  {journal} {\bibinfo
  {journal} {Int. J. Mod. Phys. A}\ }\textbf {\bibinfo {volume} {35}},\
  \bibinfo {pages} {2050075} (\bibinfo {year} {2020})},\ \Eprint
  {http://arxiv.org/abs/1807.09495} {arXiv:1807.09495 [gr-qc]} \BibitemShut
  {NoStop}%
\bibitem [{\citenamefont {Arzoumanian}\ \emph {et~al.}(2020)\citenamefont
  {Arzoumanian} \emph {et~al.}}]{Arzoumanian:2020vkk}%
  \BibitemOpen
  \bibfield  {author} {\bibinfo {author} {\bibfnamefont {Z.}~\bibnamefont
  {Arzoumanian}} \emph {et~al.} (\bibinfo {collaboration} {NANOGrav}),\ }\href
  {\doibase 10.3847/2041-8213/abd401} {\bibfield  {journal} {\bibinfo
  {journal} {Astrophys. J. Lett.}\ }\textbf {\bibinfo {volume} {905}},\
  \bibinfo {pages} {L34} (\bibinfo {year} {2020})},\ \Eprint
  {http://arxiv.org/abs/2009.04496} {arXiv:2009.04496 [astro-ph.HE]}
  \BibitemShut {NoStop}%
\bibitem [{\citenamefont {Hellings}\ and\ \citenamefont
  {Downs}(1983)}]{Hellings:1983fr}%
  \BibitemOpen
  \bibfield  {author} {\bibinfo {author} {\bibfnamefont {R.~w.}\ \bibnamefont
  {Hellings}}\ and\ \bibinfo {author} {\bibfnamefont {G.~s.}\ \bibnamefont
  {Downs}},\ }\href {\doibase 10.1086/183954} {\bibfield  {journal} {\bibinfo
  {journal} {Astrophys. J. Lett.}\ }\textbf {\bibinfo {volume} {265}},\
  \bibinfo {pages} {L39} (\bibinfo {year} {1983})}\BibitemShut {NoStop}%
\bibitem [{\citenamefont {Blasi}\ \emph {et~al.}(2021)\citenamefont {Blasi},
  \citenamefont {Brdar},\ and\ \citenamefont {Schmitz}}]{Blasi:2020mfx}%
  \BibitemOpen
  \bibfield  {author} {\bibinfo {author} {\bibfnamefont {S.}~\bibnamefont
  {Blasi}}, \bibinfo {author} {\bibfnamefont {V.}~\bibnamefont {Brdar}}, \ and\
  \bibinfo {author} {\bibfnamefont {K.}~\bibnamefont {Schmitz}},\ }\href
  {\doibase 10.1103/PhysRevLett.126.041305} {\bibfield  {journal} {\bibinfo
  {journal} {Phys. Rev. Lett.}\ }\textbf {\bibinfo {volume} {126}},\ \bibinfo
  {pages} {041305} (\bibinfo {year} {2021})},\ \Eprint
  {http://arxiv.org/abs/2009.06607} {arXiv:2009.06607 [astro-ph.CO]}
  \BibitemShut {NoStop}%
\bibitem [{\citenamefont {Ellis}\ and\ \citenamefont
  {Lewicki}(2021)}]{Ellis:2020ena}%
  \BibitemOpen
  \bibfield  {author} {\bibinfo {author} {\bibfnamefont {J.}~\bibnamefont
  {Ellis}}\ and\ \bibinfo {author} {\bibfnamefont {M.}~\bibnamefont
  {Lewicki}},\ }\href {\doibase 10.1103/PhysRevLett.126.041304} {\bibfield
  {journal} {\bibinfo  {journal} {Phys. Rev. Lett.}\ }\textbf {\bibinfo
  {volume} {126}},\ \bibinfo {pages} {041304} (\bibinfo {year} {2021})},\
  \Eprint {http://arxiv.org/abs/2009.06555} {arXiv:2009.06555 [astro-ph.CO]}
  \BibitemShut {NoStop}%
\bibitem [{\citenamefont {Buchmuller}\ \emph {et~al.}(2020)\citenamefont
  {Buchmuller}, \citenamefont {Domcke},\ and\ \citenamefont
  {Schmitz}}]{Buchmuller:2020lbh}%
  \BibitemOpen
  \bibfield  {author} {\bibinfo {author} {\bibfnamefont {W.}~\bibnamefont
  {Buchmuller}}, \bibinfo {author} {\bibfnamefont {V.}~\bibnamefont {Domcke}},
  \ and\ \bibinfo {author} {\bibfnamefont {K.}~\bibnamefont {Schmitz}},\ }\href
  {\doibase 10.1016/j.physletb.2020.135914} {\bibfield  {journal} {\bibinfo
  {journal} {Phys. Lett. B}\ }\textbf {\bibinfo {volume} {811}},\ \bibinfo
  {pages} {135914} (\bibinfo {year} {2020})},\ \Eprint
  {http://arxiv.org/abs/2009.10649} {arXiv:2009.10649 [astro-ph.CO]}
  \BibitemShut {NoStop}%
\bibitem [{\citenamefont {Samanta}\ and\ \citenamefont
  {Datta}(2021)}]{Samanta:2020cdk}%
  \BibitemOpen
  \bibfield  {author} {\bibinfo {author} {\bibfnamefont {R.}~\bibnamefont
  {Samanta}}\ and\ \bibinfo {author} {\bibfnamefont {S.}~\bibnamefont
  {Datta}},\ }\href {\doibase 10.1007/JHEP05(2021)211} {\bibfield  {journal}
  {\bibinfo  {journal} {JHEP}\ }\textbf {\bibinfo {volume} {05}},\ \bibinfo
  {pages} {211} (\bibinfo {year} {2021})},\ \Eprint
  {http://arxiv.org/abs/2009.13452} {arXiv:2009.13452 [hep-ph]} \BibitemShut
  {NoStop}%
\bibitem [{\citenamefont {Nakai}\ \emph {et~al.}(2021)\citenamefont {Nakai},
  \citenamefont {Suzuki}, \citenamefont {Takahashi},\ and\ \citenamefont
  {Yamada}}]{Nakai:2020oit}%
  \BibitemOpen
  \bibfield  {author} {\bibinfo {author} {\bibfnamefont {Y.}~\bibnamefont
  {Nakai}}, \bibinfo {author} {\bibfnamefont {M.}~\bibnamefont {Suzuki}},
  \bibinfo {author} {\bibfnamefont {F.}~\bibnamefont {Takahashi}}, \ and\
  \bibinfo {author} {\bibfnamefont {M.}~\bibnamefont {Yamada}},\ }\href
  {\doibase 10.1016/j.physletb.2021.136238} {\bibfield  {journal} {\bibinfo
  {journal} {Phys. Lett. B}\ }\textbf {\bibinfo {volume} {816}},\ \bibinfo
  {pages} {136238} (\bibinfo {year} {2021})},\ \Eprint
  {http://arxiv.org/abs/2009.09754} {arXiv:2009.09754 [astro-ph.CO]}
  \BibitemShut {NoStop}%
\bibitem [{\citenamefont {Neronov}\ \emph {et~al.}(2021)\citenamefont
  {Neronov}, \citenamefont {Pol}, \citenamefont {Caprini},\ and\ \citenamefont
  {Semikoz}}]{Neronov:2020qrl}%
  \BibitemOpen
  \bibfield  {author} {\bibinfo {author} {\bibfnamefont {A.}~\bibnamefont
  {Neronov}}, \bibinfo {author} {\bibfnamefont {A.~R.}\ \bibnamefont {Pol}},
  \bibinfo {author} {\bibfnamefont {C.}~\bibnamefont {Caprini}}, \ and\
  \bibinfo {author} {\bibfnamefont {D.}~\bibnamefont {Semikoz}},\ }\href
  {\doibase 10.1103/PhysRevD.103.L041302} {\bibfield  {journal} {\bibinfo
  {journal} {Phys. Rev. D}\ }\textbf {\bibinfo {volume} {103}},\ \bibinfo
  {pages} {L041302} (\bibinfo {year} {2021})},\ \Eprint
  {http://arxiv.org/abs/2009.14174} {arXiv:2009.14174 [astro-ph.CO]}
  \BibitemShut {NoStop}%
\bibitem [{\citenamefont {Bian}\ \emph {et~al.}(2021)\citenamefont {Bian},
  \citenamefont {Cai}, \citenamefont {Liu}, \citenamefont {Yang},\ and\
  \citenamefont {Zhou}}]{Bian:2021lmz}%
  \BibitemOpen
  \bibfield  {author} {\bibinfo {author} {\bibfnamefont {L.}~\bibnamefont
  {Bian}}, \bibinfo {author} {\bibfnamefont {R.-G.}\ \bibnamefont {Cai}},
  \bibinfo {author} {\bibfnamefont {J.}~\bibnamefont {Liu}}, \bibinfo {author}
  {\bibfnamefont {X.-Y.}\ \bibnamefont {Yang}}, \ and\ \bibinfo {author}
  {\bibfnamefont {R.}~\bibnamefont {Zhou}},\ }\href {\doibase
  10.1103/PhysRevD.103.L081301} {\bibfield  {journal} {\bibinfo  {journal}
  {Phys. Rev. D}\ }\textbf {\bibinfo {volume} {103}},\ \bibinfo {pages}
  {L081301} (\bibinfo {year} {2021})},\ \Eprint
  {http://arxiv.org/abs/2009.13893} {arXiv:2009.13893 [astro-ph.CO]}
  \BibitemShut {NoStop}%
\bibitem [{\citenamefont {Abe}\ \emph {et~al.}(2021)\citenamefont {Abe},
  \citenamefont {Tada},\ and\ \citenamefont {Ueda}}]{Abe:2020sqb}%
  \BibitemOpen
  \bibfield  {author} {\bibinfo {author} {\bibfnamefont {K.~T.}\ \bibnamefont
  {Abe}}, \bibinfo {author} {\bibfnamefont {Y.}~\bibnamefont {Tada}}, \ and\
  \bibinfo {author} {\bibfnamefont {I.}~\bibnamefont {Ueda}},\ }\href {\doibase
  10.1088/1475-7516/2021/06/048} {\bibfield  {journal} {\bibinfo  {journal}
  {JCAP}\ }\textbf {\bibinfo {volume} {06}},\ \bibinfo {pages} {048} (\bibinfo
  {year} {2021})},\ \Eprint {http://arxiv.org/abs/2010.06193} {arXiv:2010.06193
  [astro-ph.CO]} \BibitemShut {NoStop}%
\bibitem [{\citenamefont {Liu}\ \emph {et~al.}(2021)\citenamefont {Liu},
  \citenamefont {Cai},\ and\ \citenamefont {Guo}}]{Liu:2020mru}%
  \BibitemOpen
  \bibfield  {author} {\bibinfo {author} {\bibfnamefont {J.}~\bibnamefont
  {Liu}}, \bibinfo {author} {\bibfnamefont {R.-G.}\ \bibnamefont {Cai}}, \ and\
  \bibinfo {author} {\bibfnamefont {Z.-K.}\ \bibnamefont {Guo}},\ }\href
  {\doibase 10.1103/PhysRevLett.126.141303} {\bibfield  {journal} {\bibinfo
  {journal} {Phys. Rev. Lett.}\ }\textbf {\bibinfo {volume} {126}},\ \bibinfo
  {pages} {141303} (\bibinfo {year} {2021})},\ \Eprint
  {http://arxiv.org/abs/2010.03225} {arXiv:2010.03225 [astro-ph.CO]}
  \BibitemShut {NoStop}%
\bibitem [{\citenamefont {Vagnozzi}(2021)}]{Vagnozzi:2020gtf}%
  \BibitemOpen
  \bibfield  {author} {\bibinfo {author} {\bibfnamefont {S.}~\bibnamefont
  {Vagnozzi}},\ }\href {\doibase 10.1093/mnrasl/slaa203} {\bibfield  {journal}
  {\bibinfo  {journal} {Mon. Not. Roy. Astron. Soc.}\ }\textbf {\bibinfo
  {volume} {502}},\ \bibinfo {pages} {L11} (\bibinfo {year} {2021})},\ \Eprint
  {http://arxiv.org/abs/2009.13432} {arXiv:2009.13432 [astro-ph.CO]}
  \BibitemShut {NoStop}%
\bibitem [{\citenamefont {Kuroyanagi}\ \emph {et~al.}(2021)\citenamefont
  {Kuroyanagi}, \citenamefont {Takahashi},\ and\ \citenamefont
  {Yokoyama}}]{Kuroyanagi:2020sfw}%
  \BibitemOpen
  \bibfield  {author} {\bibinfo {author} {\bibfnamefont {S.}~\bibnamefont
  {Kuroyanagi}}, \bibinfo {author} {\bibfnamefont {T.}~\bibnamefont
  {Takahashi}}, \ and\ \bibinfo {author} {\bibfnamefont {S.}~\bibnamefont
  {Yokoyama}},\ }\href {\doibase 10.1088/1475-7516/2021/01/071} {\bibfield
  {journal} {\bibinfo  {journal} {JCAP}\ }\textbf {\bibinfo {volume} {01}},\
  \bibinfo {pages} {071} (\bibinfo {year} {2021})},\ \Eprint
  {http://arxiv.org/abs/2011.03323} {arXiv:2011.03323 [astro-ph.CO]}
  \BibitemShut {NoStop}%
\bibitem [{\citenamefont {Vaskonen}\ and\ \citenamefont
  {Veerm\"ae}(2021)}]{Vaskonen:2020lbd}%
  \BibitemOpen
  \bibfield  {author} {\bibinfo {author} {\bibfnamefont {V.}~\bibnamefont
  {Vaskonen}}\ and\ \bibinfo {author} {\bibfnamefont {H.}~\bibnamefont
  {Veerm\"ae}},\ }\href {\doibase 10.1103/PhysRevLett.126.051303} {\bibfield
  {journal} {\bibinfo  {journal} {Phys. Rev. Lett.}\ }\textbf {\bibinfo
  {volume} {126}},\ \bibinfo {pages} {051303} (\bibinfo {year} {2021})},\
  \Eprint {http://arxiv.org/abs/2009.07832} {arXiv:2009.07832 [astro-ph.CO]}
  \BibitemShut {NoStop}%
\bibitem [{\citenamefont {De~Luca}\ \emph {et~al.}(2021)\citenamefont
  {De~Luca}, \citenamefont {Franciolini},\ and\ \citenamefont
  {Riotto}}]{DeLuca:2020agl}%
  \BibitemOpen
  \bibfield  {author} {\bibinfo {author} {\bibfnamefont {V.}~\bibnamefont
  {De~Luca}}, \bibinfo {author} {\bibfnamefont {G.}~\bibnamefont
  {Franciolini}}, \ and\ \bibinfo {author} {\bibfnamefont {A.}~\bibnamefont
  {Riotto}},\ }\href {\doibase 10.1103/PhysRevLett.126.041303} {\bibfield
  {journal} {\bibinfo  {journal} {Phys. Rev. Lett.}\ }\textbf {\bibinfo
  {volume} {126}},\ \bibinfo {pages} {041303} (\bibinfo {year} {2021})},\
  \Eprint {http://arxiv.org/abs/2009.08268} {arXiv:2009.08268 [astro-ph.CO]}
  \BibitemShut {NoStop}%
\bibitem [{\citenamefont {Kohri}\ and\ \citenamefont
  {Terada}(2021)}]{Kohri:2020qqd}%
  \BibitemOpen
  \bibfield  {author} {\bibinfo {author} {\bibfnamefont {K.}~\bibnamefont
  {Kohri}}\ and\ \bibinfo {author} {\bibfnamefont {T.}~\bibnamefont {Terada}},\
  }\href {\doibase 10.1016/j.physletb.2020.136040} {\bibfield  {journal}
  {\bibinfo  {journal} {Phys. Lett. B}\ }\textbf {\bibinfo {volume} {813}},\
  \bibinfo {pages} {136040} (\bibinfo {year} {2021})},\ \Eprint
  {http://arxiv.org/abs/2009.11853} {arXiv:2009.11853 [astro-ph.CO]}
  \BibitemShut {NoStop}%
\bibitem [{\citenamefont {Inomata}\ \emph {et~al.}(2021)\citenamefont
  {Inomata}, \citenamefont {Kawasaki}, \citenamefont {Mukaida},\ and\
  \citenamefont {Yanagida}}]{Inomata:2020xad}%
  \BibitemOpen
  \bibfield  {author} {\bibinfo {author} {\bibfnamefont {K.}~\bibnamefont
  {Inomata}}, \bibinfo {author} {\bibfnamefont {M.}~\bibnamefont {Kawasaki}},
  \bibinfo {author} {\bibfnamefont {K.}~\bibnamefont {Mukaida}}, \ and\
  \bibinfo {author} {\bibfnamefont {T.~T.}\ \bibnamefont {Yanagida}},\ }\href
  {\doibase 10.1103/PhysRevLett.126.131301} {\bibfield  {journal} {\bibinfo
  {journal} {Phys. Rev. Lett.}\ }\textbf {\bibinfo {volume} {126}},\ \bibinfo
  {pages} {131301} (\bibinfo {year} {2021})},\ \Eprint
  {http://arxiv.org/abs/2011.01270} {arXiv:2011.01270 [astro-ph.CO]}
  \BibitemShut {NoStop}%
\bibitem [{\citenamefont {Yi}\ and\ \citenamefont {Zhu}(2021)}]{Yi:2021lxc}%
  \BibitemOpen
  \bibfield  {author} {\bibinfo {author} {\bibfnamefont {Z.}~\bibnamefont
  {Yi}}\ and\ \bibinfo {author} {\bibfnamefont {Z.-H.}\ \bibnamefont {Zhu}},\
  }\href@noop {} {\  (\bibinfo {year} {2021})},\ \Eprint
  {http://arxiv.org/abs/2105.01943} {arXiv:2105.01943 [gr-qc]} \BibitemShut
  {NoStop}%
\bibitem [{\citenamefont {Sugiyama}\ \emph {et~al.}(2021)\citenamefont
  {Sugiyama}, \citenamefont {Takhistov}, \citenamefont {Vitagliano},
  \citenamefont {Kusenko}, \citenamefont {Sasaki},\ and\ \citenamefont
  {Takada}}]{Sugiyama:2020roc}%
  \BibitemOpen
  \bibfield  {author} {\bibinfo {author} {\bibfnamefont {S.}~\bibnamefont
  {Sugiyama}}, \bibinfo {author} {\bibfnamefont {V.}~\bibnamefont {Takhistov}},
  \bibinfo {author} {\bibfnamefont {E.}~\bibnamefont {Vitagliano}}, \bibinfo
  {author} {\bibfnamefont {A.}~\bibnamefont {Kusenko}}, \bibinfo {author}
  {\bibfnamefont {M.}~\bibnamefont {Sasaki}}, \ and\ \bibinfo {author}
  {\bibfnamefont {M.}~\bibnamefont {Takada}},\ }\href {\doibase
  10.1016/j.physletb.2021.136097} {\bibfield  {journal} {\bibinfo  {journal}
  {Phys. Lett. B}\ }\textbf {\bibinfo {volume} {814}},\ \bibinfo {pages}
  {136097} (\bibinfo {year} {2021})},\ \Eprint
  {http://arxiv.org/abs/2010.02189} {arXiv:2010.02189 [astro-ph.CO]}
  \BibitemShut {NoStop}%
\bibitem [{\citenamefont {Dom\`enech}\ and\ \citenamefont
  {Pi}(2020)}]{Domenech:2020ers}%
  \BibitemOpen
  \bibfield  {author} {\bibinfo {author} {\bibfnamefont {G.}~\bibnamefont
  {Dom\`enech}}\ and\ \bibinfo {author} {\bibfnamefont {S.}~\bibnamefont
  {Pi}},\ }\href@noop {} {\  (\bibinfo {year} {2020})},\ \Eprint
  {http://arxiv.org/abs/2010.03976} {arXiv:2010.03976 [astro-ph.CO]}
  \BibitemShut {NoStop}%
\bibitem [{\citenamefont {Akrami}\ \emph {et~al.}(2020)\citenamefont {Akrami}
  \emph {et~al.}}]{Akrami:2018odb}%
  \BibitemOpen
  \bibfield  {author} {\bibinfo {author} {\bibfnamefont {Y.}~\bibnamefont
  {Akrami}} \emph {et~al.} (\bibinfo {collaboration} {Planck}),\ }\href
  {\doibase 10.1051/0004-6361/201833887} {\bibfield  {journal} {\bibinfo
  {journal} {Astron. Astrophys.}\ }\textbf {\bibinfo {volume} {641}},\ \bibinfo
  {pages} {A10} (\bibinfo {year} {2020})},\ \Eprint
  {http://arxiv.org/abs/1807.06211} {arXiv:1807.06211 [astro-ph.CO]}
  \BibitemShut {NoStop}%
\bibitem [{\citenamefont {Tegmark}\ \emph {et~al.}(2004)\citenamefont {Tegmark}
  \emph {et~al.}}]{Tegmark:2003uf}%
  \BibitemOpen
  \bibfield  {author} {\bibinfo {author} {\bibfnamefont {M.}~\bibnamefont
  {Tegmark}} \emph {et~al.} (\bibinfo {collaboration} {SDSS}),\ }\href
  {\doibase 10.1086/382125} {\bibfield  {journal} {\bibinfo  {journal}
  {Astrophys. J.}\ }\textbf {\bibinfo {volume} {606}},\ \bibinfo {pages} {702}
  (\bibinfo {year} {2004})},\ \Eprint {http://arxiv.org/abs/astro-ph/0310725}
  {arXiv:astro-ph/0310725} \BibitemShut {NoStop}%
\bibitem [{\citenamefont {Garcia-Bellido}\ and\ \citenamefont
  {Ruiz~Morales}(2017)}]{Garcia-Bellido:2017mdw}%
  \BibitemOpen
  \bibfield  {author} {\bibinfo {author} {\bibfnamefont {J.}~\bibnamefont
  {Garcia-Bellido}}\ and\ \bibinfo {author} {\bibfnamefont {E.}~\bibnamefont
  {Ruiz~Morales}},\ }\href {\doibase 10.1016/j.dark.2017.09.007} {\bibfield
  {journal} {\bibinfo  {journal} {Phys. Dark Univ.}\ }\textbf {\bibinfo
  {volume} {18}},\ \bibinfo {pages} {47} (\bibinfo {year} {2017})},\ \Eprint
  {http://arxiv.org/abs/1702.03901} {arXiv:1702.03901 [astro-ph.CO]}
  \BibitemShut {NoStop}%
\bibitem [{\citenamefont {Germani}\ and\ \citenamefont
  {Prokopec}(2017)}]{Germani:2017bcs}%
  \BibitemOpen
  \bibfield  {author} {\bibinfo {author} {\bibfnamefont {C.}~\bibnamefont
  {Germani}}\ and\ \bibinfo {author} {\bibfnamefont {T.}~\bibnamefont
  {Prokopec}},\ }\href {\doibase 10.1016/j.dark.2017.09.001} {\bibfield
  {journal} {\bibinfo  {journal} {Phys. Dark Univ.}\ }\textbf {\bibinfo
  {volume} {18}},\ \bibinfo {pages} {6} (\bibinfo {year} {2017})},\ \Eprint
  {http://arxiv.org/abs/1706.04226} {arXiv:1706.04226 [astro-ph.CO]}
  \BibitemShut {NoStop}%
\bibitem [{\citenamefont {Motohashi}\ and\ \citenamefont
  {Hu}(2017)}]{Motohashi:2017kbs}%
  \BibitemOpen
  \bibfield  {author} {\bibinfo {author} {\bibfnamefont {H.}~\bibnamefont
  {Motohashi}}\ and\ \bibinfo {author} {\bibfnamefont {W.}~\bibnamefont {Hu}},\
  }\href {\doibase 10.1103/PhysRevD.96.063503} {\bibfield  {journal} {\bibinfo
  {journal} {Phys. Rev. D}\ }\textbf {\bibinfo {volume} {96}},\ \bibinfo
  {pages} {063503} (\bibinfo {year} {2017})},\ \Eprint
  {http://arxiv.org/abs/1706.06784} {arXiv:1706.06784 [astro-ph.CO]}
  \BibitemShut {NoStop}%
\bibitem [{\citenamefont {Ezquiaga}\ \emph {et~al.}(2018)\citenamefont
  {Ezquiaga}, \citenamefont {Garcia-Bellido},\ and\ \citenamefont
  {Ruiz~Morales}}]{Ezquiaga:2017fvi}%
  \BibitemOpen
  \bibfield  {author} {\bibinfo {author} {\bibfnamefont {J.~M.}\ \bibnamefont
  {Ezquiaga}}, \bibinfo {author} {\bibfnamefont {J.}~\bibnamefont
  {Garcia-Bellido}}, \ and\ \bibinfo {author} {\bibfnamefont {E.}~\bibnamefont
  {Ruiz~Morales}},\ }\href {\doibase 10.1016/j.physletb.2017.11.039} {\bibfield
   {journal} {\bibinfo  {journal} {Phys. Lett. B}\ }\textbf {\bibinfo {volume}
  {776}},\ \bibinfo {pages} {345} (\bibinfo {year} {2018})},\ \Eprint
  {http://arxiv.org/abs/1705.04861} {arXiv:1705.04861 [astro-ph.CO]}
  \BibitemShut {NoStop}%
\bibitem [{\citenamefont {Di}\ and\ \citenamefont {Gong}(2018)}]{Gong:2017qlj}%
  \BibitemOpen
  \bibfield  {author} {\bibinfo {author} {\bibfnamefont {H.}~\bibnamefont
  {Di}}\ and\ \bibinfo {author} {\bibfnamefont {Y.}~\bibnamefont {Gong}},\
  }\href {\doibase 10.1088/1475-7516/2018/07/007} {\bibfield  {journal}
  {\bibinfo  {journal} {JCAP}\ }\textbf {\bibinfo {volume} {07}},\ \bibinfo
  {pages} {007} (\bibinfo {year} {2018})},\ \Eprint
  {http://arxiv.org/abs/1707.09578} {arXiv:1707.09578 [astro-ph.CO]}
  \BibitemShut {NoStop}%
\bibitem [{\citenamefont {Ballesteros}\ and\ \citenamefont
  {Taoso}(2018)}]{Ballesteros:2017fsr}%
  \BibitemOpen
  \bibfield  {author} {\bibinfo {author} {\bibfnamefont {G.}~\bibnamefont
  {Ballesteros}}\ and\ \bibinfo {author} {\bibfnamefont {M.}~\bibnamefont
  {Taoso}},\ }\href {\doibase 10.1103/PhysRevD.97.023501} {\bibfield  {journal}
  {\bibinfo  {journal} {Phys. Rev. D}\ }\textbf {\bibinfo {volume} {97}},\
  \bibinfo {pages} {023501} (\bibinfo {year} {2018})},\ \Eprint
  {http://arxiv.org/abs/1709.05565} {arXiv:1709.05565 [hep-ph]} \BibitemShut
  {NoStop}%
\bibitem [{\citenamefont {Dalianis}\ \emph {et~al.}(2019)\citenamefont
  {Dalianis}, \citenamefont {Kehagias},\ and\ \citenamefont
  {Tringas}}]{Dalianis:2018frf}%
  \BibitemOpen
  \bibfield  {author} {\bibinfo {author} {\bibfnamefont {I.}~\bibnamefont
  {Dalianis}}, \bibinfo {author} {\bibfnamefont {A.}~\bibnamefont {Kehagias}},
  \ and\ \bibinfo {author} {\bibfnamefont {G.}~\bibnamefont {Tringas}},\ }\href
  {\doibase 10.1088/1475-7516/2019/01/037} {\bibfield  {journal} {\bibinfo
  {journal} {JCAP}\ }\textbf {\bibinfo {volume} {01}},\ \bibinfo {pages} {037}
  (\bibinfo {year} {2019})},\ \Eprint {http://arxiv.org/abs/1805.09483}
  {arXiv:1805.09483 [astro-ph.CO]} \BibitemShut {NoStop}%
\bibitem [{\citenamefont {Gao}\ and\ \citenamefont {Guo}(2018)}]{Gao:2018pvq}%
  \BibitemOpen
  \bibfield  {author} {\bibinfo {author} {\bibfnamefont {T.-J.}\ \bibnamefont
  {Gao}}\ and\ \bibinfo {author} {\bibfnamefont {Z.-K.}\ \bibnamefont {Guo}},\
  }\href {\doibase 10.1103/PhysRevD.98.063526} {\bibfield  {journal} {\bibinfo
  {journal} {Phys. Rev. D}\ }\textbf {\bibinfo {volume} {98}},\ \bibinfo
  {pages} {063526} (\bibinfo {year} {2018})},\ \Eprint
  {http://arxiv.org/abs/1806.09320} {arXiv:1806.09320 [hep-ph]} \BibitemShut
  {NoStop}%
\bibitem [{\citenamefont {Drees}\ and\ \citenamefont
  {Xu}(2021)}]{Drees:2019xpp}%
  \BibitemOpen
  \bibfield  {author} {\bibinfo {author} {\bibfnamefont {M.}~\bibnamefont
  {Drees}}\ and\ \bibinfo {author} {\bibfnamefont {Y.}~\bibnamefont {Xu}},\
  }\href {\doibase 10.1140/epjc/s10052-021-08976-2} {\bibfield  {journal}
  {\bibinfo  {journal} {Eur. Phys. J. C}\ }\textbf {\bibinfo {volume} {81}},\
  \bibinfo {pages} {182} (\bibinfo {year} {2021})},\ \Eprint
  {http://arxiv.org/abs/1905.13581} {arXiv:1905.13581 [hep-ph]} \BibitemShut
  {NoStop}%
\bibitem [{\citenamefont {Xu}\ \emph {et~al.}(2020)\citenamefont {Xu},
  \citenamefont {Liu}, \citenamefont {Gao},\ and\ \citenamefont
  {Guo}}]{Xu:2019bdp}%
  \BibitemOpen
  \bibfield  {author} {\bibinfo {author} {\bibfnamefont {W.-T.}\ \bibnamefont
  {Xu}}, \bibinfo {author} {\bibfnamefont {J.}~\bibnamefont {Liu}}, \bibinfo
  {author} {\bibfnamefont {T.-J.}\ \bibnamefont {Gao}}, \ and\ \bibinfo
  {author} {\bibfnamefont {Z.-K.}\ \bibnamefont {Guo}},\ }\href {\doibase
  10.1103/PhysRevD.101.023505} {\bibfield  {journal} {\bibinfo  {journal}
  {Phys. Rev. D}\ }\textbf {\bibinfo {volume} {101}},\ \bibinfo {pages}
  {023505} (\bibinfo {year} {2020})},\ \Eprint
  {http://arxiv.org/abs/1907.05213} {arXiv:1907.05213 [astro-ph.CO]}
  \BibitemShut {NoStop}%
\bibitem [{\citenamefont {Fu}\ \emph {et~al.}(2019)\citenamefont {Fu},
  \citenamefont {Wu},\ and\ \citenamefont {Yu}}]{Fu:2019ttf}%
  \BibitemOpen
  \bibfield  {author} {\bibinfo {author} {\bibfnamefont {C.}~\bibnamefont
  {Fu}}, \bibinfo {author} {\bibfnamefont {P.}~\bibnamefont {Wu}}, \ and\
  \bibinfo {author} {\bibfnamefont {H.}~\bibnamefont {Yu}},\ }\href {\doibase
  10.1103/PhysRevD.100.063532} {\bibfield  {journal} {\bibinfo  {journal}
  {Phys. Rev. D}\ }\textbf {\bibinfo {volume} {100}},\ \bibinfo {pages}
  {063532} (\bibinfo {year} {2019})},\ \Eprint
  {http://arxiv.org/abs/1907.05042} {arXiv:1907.05042 [astro-ph.CO]}
  \BibitemShut {NoStop}%
\bibitem [{\citenamefont {Lin}\ \emph {et~al.}(2020)\citenamefont {Lin},
  \citenamefont {Gao}, \citenamefont {Gong}, \citenamefont {Lu}, \citenamefont
  {Zhang},\ and\ \citenamefont {Zhang}}]{Lin:2020goi}%
  \BibitemOpen
  \bibfield  {author} {\bibinfo {author} {\bibfnamefont {J.}~\bibnamefont
  {Lin}}, \bibinfo {author} {\bibfnamefont {Q.}~\bibnamefont {Gao}}, \bibinfo
  {author} {\bibfnamefont {Y.}~\bibnamefont {Gong}}, \bibinfo {author}
  {\bibfnamefont {Y.}~\bibnamefont {Lu}}, \bibinfo {author} {\bibfnamefont
  {C.}~\bibnamefont {Zhang}}, \ and\ \bibinfo {author} {\bibfnamefont
  {F.}~\bibnamefont {Zhang}},\ }\href {\doibase 10.1103/PhysRevD.101.103515}
  {\bibfield  {journal} {\bibinfo  {journal} {Phys. Rev. D}\ }\textbf {\bibinfo
  {volume} {101}},\ \bibinfo {pages} {103515} (\bibinfo {year} {2020})},\
  \Eprint {http://arxiv.org/abs/2001.05909} {arXiv:2001.05909 [gr-qc]}
  \BibitemShut {NoStop}%
\bibitem [{\citenamefont {Fu}\ \emph {et~al.}(2020{\natexlab{b}})\citenamefont
  {Fu}, \citenamefont {Wu},\ and\ \citenamefont {Yu}}]{Fu:2020lob}%
  \BibitemOpen
  \bibfield  {author} {\bibinfo {author} {\bibfnamefont {C.}~\bibnamefont
  {Fu}}, \bibinfo {author} {\bibfnamefont {P.}~\bibnamefont {Wu}}, \ and\
  \bibinfo {author} {\bibfnamefont {H.}~\bibnamefont {Yu}},\ }\href {\doibase
  10.1103/PhysRevD.102.043527} {\bibfield  {journal} {\bibinfo  {journal}
  {Phys. Rev. D}\ }\textbf {\bibinfo {volume} {102}},\ \bibinfo {pages}
  {043527} (\bibinfo {year} {2020}{\natexlab{b}})},\ \Eprint
  {http://arxiv.org/abs/2006.03768} {arXiv:2006.03768 [astro-ph.CO]}
  \BibitemShut {NoStop}%
\bibitem [{\citenamefont {Yi}\ \emph {et~al.}(2021)\citenamefont {Yi},
  \citenamefont {Gao}, \citenamefont {Gong},\ and\ \citenamefont
  {Zhu}}]{Yi:2020cut}%
  \BibitemOpen
  \bibfield  {author} {\bibinfo {author} {\bibfnamefont {Z.}~\bibnamefont
  {Yi}}, \bibinfo {author} {\bibfnamefont {Q.}~\bibnamefont {Gao}}, \bibinfo
  {author} {\bibfnamefont {Y.}~\bibnamefont {Gong}}, \ and\ \bibinfo {author}
  {\bibfnamefont {Z.-h.}\ \bibnamefont {Zhu}},\ }\href {\doibase
  10.1103/PhysRevD.103.063534} {\bibfield  {journal} {\bibinfo  {journal}
  {Phys. Rev. D}\ }\textbf {\bibinfo {volume} {103}},\ \bibinfo {pages}
  {063534} (\bibinfo {year} {2021})},\ \Eprint
  {http://arxiv.org/abs/2011.10606} {arXiv:2011.10606 [astro-ph.CO]}
  \BibitemShut {NoStop}%
\bibitem [{\citenamefont {Ballesteros}\ \emph {et~al.}(2019)\citenamefont
  {Ballesteros}, \citenamefont {Beltran~Jimenez},\ and\ \citenamefont
  {Pieroni}}]{Ballesteros:2018wlw}%
  \BibitemOpen
  \bibfield  {author} {\bibinfo {author} {\bibfnamefont {G.}~\bibnamefont
  {Ballesteros}}, \bibinfo {author} {\bibfnamefont {J.}~\bibnamefont
  {Beltran~Jimenez}}, \ and\ \bibinfo {author} {\bibfnamefont {M.}~\bibnamefont
  {Pieroni}},\ }\href {\doibase 10.1088/1475-7516/2019/06/016} {\bibfield
  {journal} {\bibinfo  {journal} {JCAP}\ }\textbf {\bibinfo {volume} {06}},\
  \bibinfo {pages} {016} (\bibinfo {year} {2019})},\ \Eprint
  {http://arxiv.org/abs/1811.03065} {arXiv:1811.03065 [astro-ph.CO]}
  \BibitemShut {NoStop}%
\bibitem [{\citenamefont {Kamenshchik}\ \emph {et~al.}(2019)\citenamefont
  {Kamenshchik}, \citenamefont {Tronconi}, \citenamefont {Vardanyan},\ and\
  \citenamefont {Venturi}}]{Kamenshchik:2018sig}%
  \BibitemOpen
  \bibfield  {author} {\bibinfo {author} {\bibfnamefont {A.~Y.}\ \bibnamefont
  {Kamenshchik}}, \bibinfo {author} {\bibfnamefont {A.}~\bibnamefont
  {Tronconi}}, \bibinfo {author} {\bibfnamefont {T.}~\bibnamefont {Vardanyan}},
  \ and\ \bibinfo {author} {\bibfnamefont {G.}~\bibnamefont {Venturi}},\ }\href
  {\doibase 10.1016/j.physletb.2019.02.036} {\bibfield  {journal} {\bibinfo
  {journal} {Phys. Lett. B}\ }\textbf {\bibinfo {volume} {791}},\ \bibinfo
  {pages} {201} (\bibinfo {year} {2019})},\ \Eprint
  {http://arxiv.org/abs/1812.02547} {arXiv:1812.02547 [gr-qc]} \BibitemShut
  {NoStop}%
\bibitem [{\citenamefont {Cai}\ \emph {et~al.}(2018)\citenamefont {Cai},
  \citenamefont {Tong}, \citenamefont {Wang},\ and\ \citenamefont
  {Yan}}]{Cai:2018tuh}%
  \BibitemOpen
  \bibfield  {author} {\bibinfo {author} {\bibfnamefont {Y.-F.}\ \bibnamefont
  {Cai}}, \bibinfo {author} {\bibfnamefont {X.}~\bibnamefont {Tong}}, \bibinfo
  {author} {\bibfnamefont {D.-G.}\ \bibnamefont {Wang}}, \ and\ \bibinfo
  {author} {\bibfnamefont {S.-F.}\ \bibnamefont {Yan}},\ }\href {\doibase
  10.1103/PhysRevLett.121.081306} {\bibfield  {journal} {\bibinfo  {journal}
  {Phys. Rev. Lett.}\ }\textbf {\bibinfo {volume} {121}},\ \bibinfo {pages}
  {081306} (\bibinfo {year} {2018})},\ \Eprint
  {http://arxiv.org/abs/1805.03639} {arXiv:1805.03639 [astro-ph.CO]}
  \BibitemShut {NoStop}%
\bibitem [{\citenamefont {Chen}\ and\ \citenamefont
  {Cai}(2019)}]{Chen:2019zza}%
  \BibitemOpen
  \bibfield  {author} {\bibinfo {author} {\bibfnamefont {C.}~\bibnamefont
  {Chen}}\ and\ \bibinfo {author} {\bibfnamefont {Y.-F.}\ \bibnamefont {Cai}},\
  }\href {\doibase 10.1088/1475-7516/2019/10/068} {\bibfield  {journal}
  {\bibinfo  {journal} {JCAP}\ }\textbf {\bibinfo {volume} {10}},\ \bibinfo
  {pages} {068} (\bibinfo {year} {2019})},\ \Eprint
  {http://arxiv.org/abs/1908.03942} {arXiv:1908.03942 [astro-ph.CO]}
  \BibitemShut {NoStop}%
\bibitem [{\citenamefont {Chen}\ \emph {et~al.}(2020)\citenamefont {Chen},
  \citenamefont {Ma},\ and\ \citenamefont {Cai}}]{Chen:2020uhe}%
  \BibitemOpen
  \bibfield  {author} {\bibinfo {author} {\bibfnamefont {C.}~\bibnamefont
  {Chen}}, \bibinfo {author} {\bibfnamefont {X.-H.}\ \bibnamefont {Ma}}, \ and\
  \bibinfo {author} {\bibfnamefont {Y.-F.}\ \bibnamefont {Cai}},\ }\href
  {\doibase 10.1103/PhysRevD.102.063526} {\bibfield  {journal} {\bibinfo
  {journal} {Phys. Rev. D}\ }\textbf {\bibinfo {volume} {102}},\ \bibinfo
  {pages} {063526} (\bibinfo {year} {2020})},\ \Eprint
  {http://arxiv.org/abs/2003.03821} {arXiv:2003.03821 [astro-ph.CO]}
  \BibitemShut {NoStop}%
\bibitem [{\citenamefont {Cai}\ \emph {et~al.}(2020)\citenamefont {Cai},
  \citenamefont {Guo}, \citenamefont {Liu}, \citenamefont {Liu},\ and\
  \citenamefont {Yang}}]{Cai:2019bmk}%
  \BibitemOpen
  \bibfield  {author} {\bibinfo {author} {\bibfnamefont {R.-G.}\ \bibnamefont
  {Cai}}, \bibinfo {author} {\bibfnamefont {Z.-K.}\ \bibnamefont {Guo}},
  \bibinfo {author} {\bibfnamefont {J.}~\bibnamefont {Liu}}, \bibinfo {author}
  {\bibfnamefont {L.}~\bibnamefont {Liu}}, \ and\ \bibinfo {author}
  {\bibfnamefont {X.-Y.}\ \bibnamefont {Yang}},\ }\href {\doibase
  10.1088/1475-7516/2020/06/013} {\bibfield  {journal} {\bibinfo  {journal}
  {JCAP}\ }\textbf {\bibinfo {volume} {06}},\ \bibinfo {pages} {013} (\bibinfo
  {year} {2020})},\ \Eprint {http://arxiv.org/abs/1912.10437} {arXiv:1912.10437
  [astro-ph.CO]} \BibitemShut {NoStop}%
\bibitem [{\citenamefont {Zhou}\ \emph {et~al.}(2020)\citenamefont {Zhou},
  \citenamefont {Jiang}, \citenamefont {Cai}, \citenamefont {Sasaki},\ and\
  \citenamefont {Pi}}]{Zhou:2020kkf}%
  \BibitemOpen
  \bibfield  {author} {\bibinfo {author} {\bibfnamefont {Z.}~\bibnamefont
  {Zhou}}, \bibinfo {author} {\bibfnamefont {J.}~\bibnamefont {Jiang}},
  \bibinfo {author} {\bibfnamefont {Y.-F.}\ \bibnamefont {Cai}}, \bibinfo
  {author} {\bibfnamefont {M.}~\bibnamefont {Sasaki}}, \ and\ \bibinfo {author}
  {\bibfnamefont {S.}~\bibnamefont {Pi}},\ }\href {\doibase
  10.1103/PhysRevD.102.103527} {\bibfield  {journal} {\bibinfo  {journal}
  {Phys. Rev. D}\ }\textbf {\bibinfo {volume} {102}},\ \bibinfo {pages}
  {103527} (\bibinfo {year} {2020})},\ \Eprint
  {http://arxiv.org/abs/2010.03537} {arXiv:2010.03537 [astro-ph.CO]}
  \BibitemShut {NoStop}%
\bibitem [{\citenamefont {Liu}\ and\ \citenamefont {Xu}(2021)}]{Liu:2021rgq}%
  \BibitemOpen
  \bibfield  {author} {\bibinfo {author} {\bibfnamefont {L.-H.}\ \bibnamefont
  {Liu}}\ and\ \bibinfo {author} {\bibfnamefont {W.-L.}\ \bibnamefont {Xu}},\
  }\href@noop {} {\  (\bibinfo {year} {2021})},\ \Eprint
  {http://arxiv.org/abs/2107.07310} {arXiv:2107.07310 [astro-ph.CO]}
  \BibitemShut {NoStop}%
\bibitem [{\citenamefont {Ashoorioon}\ \emph {et~al.}(2021)\citenamefont
  {Ashoorioon}, \citenamefont {Rostami},\ and\ \citenamefont
  {Firouzjaee}}]{Ashoorioon:2019xqc}%
  \BibitemOpen
  \bibfield  {author} {\bibinfo {author} {\bibfnamefont {A.}~\bibnamefont
  {Ashoorioon}}, \bibinfo {author} {\bibfnamefont {A.}~\bibnamefont {Rostami}},
  \ and\ \bibinfo {author} {\bibfnamefont {J.~T.}\ \bibnamefont {Firouzjaee}},\
  }\href {\doibase 10.1007/JHEP07(2021)087} {\bibfield  {journal} {\bibinfo
  {journal} {JHEP}\ }\textbf {\bibinfo {volume} {07}},\ \bibinfo {pages} {087}
  (\bibinfo {year} {2021})},\ \Eprint {http://arxiv.org/abs/1912.13326}
  {arXiv:1912.13326 [astro-ph.CO]} \BibitemShut {NoStop}%
\bibitem [{\citenamefont {Cheng}\ \emph {et~al.}(2018)\citenamefont {Cheng},
  \citenamefont {Lee},\ and\ \citenamefont {Ng}}]{Cheng:2018yyr}%
  \BibitemOpen
  \bibfield  {author} {\bibinfo {author} {\bibfnamefont {S.-L.}\ \bibnamefont
  {Cheng}}, \bibinfo {author} {\bibfnamefont {W.}~\bibnamefont {Lee}}, \ and\
  \bibinfo {author} {\bibfnamefont {K.-W.}\ \bibnamefont {Ng}},\ }\href
  {\doibase 10.1088/1475-7516/2018/07/001} {\bibfield  {journal} {\bibinfo
  {journal} {JCAP}\ }\textbf {\bibinfo {volume} {07}},\ \bibinfo {pages} {001}
  (\bibinfo {year} {2018})},\ \Eprint {http://arxiv.org/abs/1801.09050}
  {arXiv:1801.09050 [astro-ph.CO]} \BibitemShut {NoStop}%
\bibitem [{\citenamefont {\"Ozsoy}\ and\ \citenamefont
  {Lalak}(2021)}]{Ozsoy:2020kat}%
  \BibitemOpen
  \bibfield  {author} {\bibinfo {author} {\bibfnamefont {O.}~\bibnamefont
  {\"Ozsoy}}\ and\ \bibinfo {author} {\bibfnamefont {Z.}~\bibnamefont
  {Lalak}},\ }\href {\doibase 10.1088/1475-7516/2021/01/040} {\bibfield
  {journal} {\bibinfo  {journal} {JCAP}\ }\textbf {\bibinfo {volume} {01}},\
  \bibinfo {pages} {040} (\bibinfo {year} {2021})},\ \Eprint
  {http://arxiv.org/abs/2008.07549} {arXiv:2008.07549 [astro-ph.CO]}
  \BibitemShut {NoStop}%
\bibitem [{\citenamefont {Cook}\ and\ \citenamefont
  {Sorbo}(2012)}]{Cook:2011hg}%
  \BibitemOpen
  \bibfield  {author} {\bibinfo {author} {\bibfnamefont {J.~L.}\ \bibnamefont
  {Cook}}\ and\ \bibinfo {author} {\bibfnamefont {L.}~\bibnamefont {Sorbo}},\
  }\href {\doibase 10.1103/PhysRevD.85.023534} {\bibfield  {journal} {\bibinfo
  {journal} {Phys. Rev. D}\ }\textbf {\bibinfo {volume} {85}},\ \bibinfo
  {pages} {023534} (\bibinfo {year} {2012})},\ \bibinfo {note} {[Erratum:
  Phys.Rev.D 86, 069901(E) (2012)]},\ \Eprint {http://arxiv.org/abs/1109.0022}
  {arXiv:1109.0022 [astro-ph.CO]} \BibitemShut {NoStop}%
\bibitem [{\citenamefont {Goolsby-Cole}\ and\ \citenamefont
  {Sorbo}(2017)}]{Goolsby-Cole:2017hod}%
  \BibitemOpen
  \bibfield  {author} {\bibinfo {author} {\bibfnamefont {C.}~\bibnamefont
  {Goolsby-Cole}}\ and\ \bibinfo {author} {\bibfnamefont {L.}~\bibnamefont
  {Sorbo}},\ }\href {\doibase 10.1088/1475-7516/2017/08/005} {\bibfield
  {journal} {\bibinfo  {journal} {JCAP}\ }\textbf {\bibinfo {volume} {08}},\
  \bibinfo {pages} {005} (\bibinfo {year} {2017})},\ \Eprint
  {http://arxiv.org/abs/1705.03755} {arXiv:1705.03755 [astro-ph.CO]}
  \BibitemShut {NoStop}%
\bibitem [{\citenamefont {Biagetti}\ \emph {et~al.}(2013)\citenamefont
  {Biagetti}, \citenamefont {Fasiello},\ and\ \citenamefont
  {Riotto}}]{Biagetti:2013kwa}%
  \BibitemOpen
  \bibfield  {author} {\bibinfo {author} {\bibfnamefont {M.}~\bibnamefont
  {Biagetti}}, \bibinfo {author} {\bibfnamefont {M.}~\bibnamefont {Fasiello}},
  \ and\ \bibinfo {author} {\bibfnamefont {A.}~\bibnamefont {Riotto}},\ }\href
  {\doibase 10.1103/PhysRevD.88.103518} {\bibfield  {journal} {\bibinfo
  {journal} {Phys. Rev. D}\ }\textbf {\bibinfo {volume} {88}},\ \bibinfo
  {pages} {103518} (\bibinfo {year} {2013})},\ \Eprint
  {http://arxiv.org/abs/1305.7241} {arXiv:1305.7241 [astro-ph.CO]} \BibitemShut
  {NoStop}%
\bibitem [{\citenamefont {Biagetti}\ \emph {et~al.}(2015)\citenamefont
  {Biagetti}, \citenamefont {Dimastrogiovanni}, \citenamefont {Fasiello},\ and\
  \citenamefont {Peloso}}]{Biagetti:2014asa}%
  \BibitemOpen
  \bibfield  {author} {\bibinfo {author} {\bibfnamefont {M.}~\bibnamefont
  {Biagetti}}, \bibinfo {author} {\bibfnamefont {E.}~\bibnamefont
  {Dimastrogiovanni}}, \bibinfo {author} {\bibfnamefont {M.}~\bibnamefont
  {Fasiello}}, \ and\ \bibinfo {author} {\bibfnamefont {M.}~\bibnamefont
  {Peloso}},\ }\href {\doibase 10.1088/1475-7516/2015/04/011} {\bibfield
  {journal} {\bibinfo  {journal} {JCAP}\ }\textbf {\bibinfo {volume} {04}},\
  \bibinfo {pages} {011} (\bibinfo {year} {2015})},\ \Eprint
  {http://arxiv.org/abs/1411.3029} {arXiv:1411.3029 [astro-ph.CO]} \BibitemShut
  {NoStop}%
\bibitem [{\citenamefont {Fujita}\ \emph {et~al.}(2015)\citenamefont {Fujita},
  \citenamefont {Yokoyama},\ and\ \citenamefont {Yokoyama}}]{Fujita:2014oba}%
  \BibitemOpen
  \bibfield  {author} {\bibinfo {author} {\bibfnamefont {T.}~\bibnamefont
  {Fujita}}, \bibinfo {author} {\bibfnamefont {J.}~\bibnamefont {Yokoyama}}, \
  and\ \bibinfo {author} {\bibfnamefont {S.}~\bibnamefont {Yokoyama}},\ }\href
  {\doibase 10.1093/ptep/ptv037} {\bibfield  {journal} {\bibinfo  {journal}
  {PTEP}\ }\textbf {\bibinfo {volume} {2015}},\ \bibinfo {pages} {43E01}
  (\bibinfo {year} {2015})},\ \Eprint {http://arxiv.org/abs/1411.3658}
  {arXiv:1411.3658 [astro-ph.CO]} \BibitemShut {NoStop}%
\bibitem [{\citenamefont {Starobinsky}\ \emph {et~al.}(2001)\citenamefont
  {Starobinsky}, \citenamefont {Tsujikawa},\ and\ \citenamefont
  {Yokoyama}}]{Starobinsky:2001xq}%
  \BibitemOpen
  \bibfield  {author} {\bibinfo {author} {\bibfnamefont {A.~A.}\ \bibnamefont
  {Starobinsky}}, \bibinfo {author} {\bibfnamefont {S.}~\bibnamefont
  {Tsujikawa}}, \ and\ \bibinfo {author} {\bibfnamefont {J.}~\bibnamefont
  {Yokoyama}},\ }\href {\doibase 10.1016/S0550-3213(01)00322-4} {\bibfield
  {journal} {\bibinfo  {journal} {Nucl. Phys. B}\ }\textbf {\bibinfo {volume}
  {610}},\ \bibinfo {pages} {383} (\bibinfo {year} {2001})},\ \Eprint
  {http://arxiv.org/abs/astro-ph/0107555} {arXiv:astro-ph/0107555} \BibitemShut
  {NoStop}%
\bibitem [{\citenamefont {Dom\`enech}\ and\ \citenamefont
  {Sasaki}(2021)}]{Domenech:2020xin}%
  \BibitemOpen
  \bibfield  {author} {\bibinfo {author} {\bibfnamefont {G.}~\bibnamefont
  {Dom\`enech}}\ and\ \bibinfo {author} {\bibfnamefont {M.}~\bibnamefont
  {Sasaki}},\ }\href {\doibase 10.1103/PhysRevD.103.063531} {\bibfield
  {journal} {\bibinfo  {journal} {Phys. Rev. D}\ }\textbf {\bibinfo {volume}
  {103}},\ \bibinfo {pages} {063531} (\bibinfo {year} {2021})},\ \Eprint
  {http://arxiv.org/abs/2012.14016} {arXiv:2012.14016 [gr-qc]} \BibitemShut
  {NoStop}%
\bibitem [{\citenamefont {van~de Bruck}\ and\ \citenamefont
  {Robinson}(2014)}]{vandeBruck:2014ata}%
  \BibitemOpen
  \bibfield  {author} {\bibinfo {author} {\bibfnamefont {C.}~\bibnamefont
  {van~de Bruck}}\ and\ \bibinfo {author} {\bibfnamefont {M.}~\bibnamefont
  {Robinson}},\ }\href {\doibase 10.1088/1475-7516/2014/08/024} {\bibfield
  {journal} {\bibinfo  {journal} {JCAP}\ }\textbf {\bibinfo {volume} {08}},\
  \bibinfo {pages} {024} (\bibinfo {year} {2014})},\ \Eprint
  {http://arxiv.org/abs/1404.7806} {arXiv:1404.7806 [astro-ph.CO]} \BibitemShut
  {NoStop}%
\bibitem [{\citenamefont {Gordon}\ \emph {et~al.}(2000)\citenamefont {Gordon},
  \citenamefont {Wands}, \citenamefont {Bassett},\ and\ \citenamefont
  {Maartens}}]{Gordon:2000hv}%
  \BibitemOpen
  \bibfield  {author} {\bibinfo {author} {\bibfnamefont {C.}~\bibnamefont
  {Gordon}}, \bibinfo {author} {\bibfnamefont {D.}~\bibnamefont {Wands}},
  \bibinfo {author} {\bibfnamefont {B.~A.}\ \bibnamefont {Bassett}}, \ and\
  \bibinfo {author} {\bibfnamefont {R.}~\bibnamefont {Maartens}},\ }\href
  {\doibase 10.1103/PhysRevD.63.023506} {\bibfield  {journal} {\bibinfo
  {journal} {Phys. Rev. D}\ }\textbf {\bibinfo {volume} {63}},\ \bibinfo
  {pages} {023506} (\bibinfo {year} {2000})},\ \Eprint
  {http://arxiv.org/abs/astro-ph/0009131} {arXiv:astro-ph/0009131} \BibitemShut
  {NoStop}%
\bibitem [{\citenamefont {Ando}\ \emph {et~al.}(2018)\citenamefont {Ando},
  \citenamefont {Inomata},\ and\ \citenamefont {Kawasaki}}]{Ando:2018qdb}%
  \BibitemOpen
  \bibfield  {author} {\bibinfo {author} {\bibfnamefont {K.}~\bibnamefont
  {Ando}}, \bibinfo {author} {\bibfnamefont {K.}~\bibnamefont {Inomata}}, \
  and\ \bibinfo {author} {\bibfnamefont {M.}~\bibnamefont {Kawasaki}},\ }\href
  {\doibase 10.1103/PhysRevD.97.103528} {\bibfield  {journal} {\bibinfo
  {journal} {Phys. Rev. D}\ }\textbf {\bibinfo {volume} {97}},\ \bibinfo
  {pages} {103528} (\bibinfo {year} {2018})},\ \Eprint
  {http://arxiv.org/abs/1802.06393} {arXiv:1802.06393 [astro-ph.CO]}
  \BibitemShut {NoStop}%
\bibitem [{\citenamefont {Musco}\ \emph {et~al.}(2005)\citenamefont {Musco},
  \citenamefont {Miller},\ and\ \citenamefont {Rezzolla}}]{Musco:2004ak}%
  \BibitemOpen
  \bibfield  {author} {\bibinfo {author} {\bibfnamefont {I.}~\bibnamefont
  {Musco}}, \bibinfo {author} {\bibfnamefont {J.~C.}\ \bibnamefont {Miller}}, \
  and\ \bibinfo {author} {\bibfnamefont {L.}~\bibnamefont {Rezzolla}},\ }\href
  {\doibase 10.1088/0264-9381/22/7/013} {\bibfield  {journal} {\bibinfo
  {journal} {Class. Quant. Grav.}\ }\textbf {\bibinfo {volume} {22}},\ \bibinfo
  {pages} {1405} (\bibinfo {year} {2005})},\ \Eprint
  {http://arxiv.org/abs/gr-qc/0412063} {arXiv:gr-qc/0412063} \BibitemShut
  {NoStop}%
\bibitem [{\citenamefont {Blais}\ \emph {et~al.}(2003)\citenamefont {Blais},
  \citenamefont {Bringmann}, \citenamefont {Kiefer},\ and\ \citenamefont
  {Polarski}}]{Blais:2002gw}%
  \BibitemOpen
  \bibfield  {author} {\bibinfo {author} {\bibfnamefont {D.}~\bibnamefont
  {Blais}}, \bibinfo {author} {\bibfnamefont {T.}~\bibnamefont {Bringmann}},
  \bibinfo {author} {\bibfnamefont {C.}~\bibnamefont {Kiefer}}, \ and\ \bibinfo
  {author} {\bibfnamefont {D.}~\bibnamefont {Polarski}},\ }\href {\doibase
  10.1103/PhysRevD.67.024024} {\bibfield  {journal} {\bibinfo  {journal} {Phys.
  Rev. D}\ }\textbf {\bibinfo {volume} {67}},\ \bibinfo {pages} {024024}
  (\bibinfo {year} {2003})},\ \Eprint {http://arxiv.org/abs/astro-ph/0206262}
  {arXiv:astro-ph/0206262} \BibitemShut {NoStop}%
\bibitem [{\citenamefont {Josan}\ \emph {et~al.}(2009)\citenamefont {Josan},
  \citenamefont {Green},\ and\ \citenamefont {Malik}}]{Josan:2009qn}%
  \BibitemOpen
  \bibfield  {author} {\bibinfo {author} {\bibfnamefont {A.~S.}\ \bibnamefont
  {Josan}}, \bibinfo {author} {\bibfnamefont {A.~M.}\ \bibnamefont {Green}}, \
  and\ \bibinfo {author} {\bibfnamefont {K.~A.}\ \bibnamefont {Malik}},\ }\href
  {\doibase 10.1103/PhysRevD.79.103520} {\bibfield  {journal} {\bibinfo
  {journal} {Phys. Rev. D}\ }\textbf {\bibinfo {volume} {79}},\ \bibinfo
  {pages} {103520} (\bibinfo {year} {2009})},\ \Eprint
  {http://arxiv.org/abs/0903.3184} {arXiv:0903.3184 [astro-ph.CO]} \BibitemShut
  {NoStop}%
\bibitem [{\citenamefont {Sasaki}\ \emph {et~al.}(2018)\citenamefont {Sasaki},
  \citenamefont {Suyama}, \citenamefont {Tanaka},\ and\ \citenamefont
  {Yokoyama}}]{Sasaki:2018dmp}%
  \BibitemOpen
  \bibfield  {author} {\bibinfo {author} {\bibfnamefont {M.}~\bibnamefont
  {Sasaki}}, \bibinfo {author} {\bibfnamefont {T.}~\bibnamefont {Suyama}},
  \bibinfo {author} {\bibfnamefont {T.}~\bibnamefont {Tanaka}}, \ and\ \bibinfo
  {author} {\bibfnamefont {S.}~\bibnamefont {Yokoyama}},\ }\href {\doibase
  10.1088/1361-6382/aaa7b4} {\bibfield  {journal} {\bibinfo  {journal} {Class.
  Quant. Grav.}\ }\textbf {\bibinfo {volume} {35}},\ \bibinfo {pages} {063001}
  (\bibinfo {year} {2018})},\ \Eprint {http://arxiv.org/abs/1801.05235}
  {arXiv:1801.05235 [astro-ph.CO]} \BibitemShut {NoStop}%
\bibitem [{\citenamefont {Zhao}\ and\ \citenamefont
  {Zhang}(2006)}]{Zhao:2006mm}%
  \BibitemOpen
  \bibfield  {author} {\bibinfo {author} {\bibfnamefont {W.}~\bibnamefont
  {Zhao}}\ and\ \bibinfo {author} {\bibfnamefont {Y.}~\bibnamefont {Zhang}},\
  }\href {\doibase 10.1103/PhysRevD.74.043503} {\bibfield  {journal} {\bibinfo
  {journal} {Phys. Rev. D}\ }\textbf {\bibinfo {volume} {74}},\ \bibinfo
  {pages} {043503} (\bibinfo {year} {2006})},\ \Eprint
  {http://arxiv.org/abs/astro-ph/0604458} {arXiv:astro-ph/0604458} \BibitemShut
  {NoStop}%
\bibitem [{\citenamefont {Mather}\ \emph {et~al.}(1994)\citenamefont {Mather}
  \emph {et~al.}}]{Mather:1993ij}%
  \BibitemOpen
  \bibfield  {author} {\bibinfo {author} {\bibfnamefont {J.~C.}\ \bibnamefont
  {Mather}} \emph {et~al.},\ }\href {\doibase 10.1086/173574} {\bibfield
  {journal} {\bibinfo  {journal} {Astrophys. J.}\ }\textbf {\bibinfo {volume}
  {420}},\ \bibinfo {pages} {439} (\bibinfo {year} {1994})}\BibitemShut
  {NoStop}%
\bibitem [{\citenamefont {Fixsen}\ \emph {et~al.}(1996)\citenamefont {Fixsen},
  \citenamefont {Cheng}, \citenamefont {Gales}, \citenamefont {Mather},
  \citenamefont {Shafer},\ and\ \citenamefont {Wright}}]{Fixsen:1996nj}%
  \BibitemOpen
  \bibfield  {author} {\bibinfo {author} {\bibfnamefont {D.~J.}\ \bibnamefont
  {Fixsen}}, \bibinfo {author} {\bibfnamefont {E.~S.}\ \bibnamefont {Cheng}},
  \bibinfo {author} {\bibfnamefont {J.~M.}\ \bibnamefont {Gales}}, \bibinfo
  {author} {\bibfnamefont {J.~C.}\ \bibnamefont {Mather}}, \bibinfo {author}
  {\bibfnamefont {R.~A.}\ \bibnamefont {Shafer}}, \ and\ \bibinfo {author}
  {\bibfnamefont {E.~L.}\ \bibnamefont {Wright}},\ }\href {\doibase
  10.1086/178173} {\bibfield  {journal} {\bibinfo  {journal} {Astrophys. J.}\
  }\textbf {\bibinfo {volume} {473}},\ \bibinfo {pages} {576} (\bibinfo {year}
  {1996})},\ \Eprint {http://arxiv.org/abs/astro-ph/9605054}
  {arXiv:astro-ph/9605054} \BibitemShut {NoStop}%
\bibitem [{\citenamefont {Kogut}\ \emph {et~al.}(2011)\citenamefont {Kogut}
  \emph {et~al.}}]{Kogut:2011xw}%
  \BibitemOpen
  \bibfield  {author} {\bibinfo {author} {\bibfnamefont {A.}~\bibnamefont
  {Kogut}} \emph {et~al.},\ }\href {\doibase 10.1088/1475-7516/2011/07/025}
  {\bibfield  {journal} {\bibinfo  {journal} {JCAP}\ }\textbf {\bibinfo
  {volume} {07}},\ \bibinfo {pages} {025} (\bibinfo {year} {2011})},\ \Eprint
  {http://arxiv.org/abs/1105.2044} {arXiv:1105.2044 [astro-ph.CO]} \BibitemShut
  {NoStop}%
\bibitem [{\citenamefont {Chluba}\ \emph {et~al.}(2019)\citenamefont {Chluba}
  \emph {et~al.}}]{Chluba:2019nxa}%
  \BibitemOpen
  \bibfield  {author} {\bibinfo {author} {\bibfnamefont {J.}~\bibnamefont
  {Chluba}} \emph {et~al.},\ }\href {\doibase 10.1007/s10686-021-09729-5} {\
  (\bibinfo {year} {2019}),\ 10.1007/s10686-021-09729-5},\ \Eprint
  {http://arxiv.org/abs/1909.01593} {arXiv:1909.01593 [astro-ph.CO]}
  \BibitemShut {NoStop}%
\bibitem [{\citenamefont {Dasgupta}\ \emph {et~al.}(2020)\citenamefont
  {Dasgupta}, \citenamefont {Laha},\ and\ \citenamefont
  {Ray}}]{Dasgupta:2019cae}%
  \BibitemOpen
  \bibfield  {author} {\bibinfo {author} {\bibfnamefont {B.}~\bibnamefont
  {Dasgupta}}, \bibinfo {author} {\bibfnamefont {R.}~\bibnamefont {Laha}}, \
  and\ \bibinfo {author} {\bibfnamefont {A.}~\bibnamefont {Ray}},\ }\href
  {\doibase 10.1103/PhysRevLett.125.101101} {\bibfield  {journal} {\bibinfo
  {journal} {Phys. Rev. Lett.}\ }\textbf {\bibinfo {volume} {125}},\ \bibinfo
  {pages} {101101} (\bibinfo {year} {2020})},\ \Eprint
  {http://arxiv.org/abs/1912.01014} {arXiv:1912.01014 [hep-ph]} \BibitemShut
  {NoStop}%
\bibitem [{\citenamefont {Carr}\ \emph {et~al.}(2010)\citenamefont {Carr},
  \citenamefont {Kohri}, \citenamefont {Sendouda},\ and\ \citenamefont
  {Yokoyama}}]{Carr:2009jm}%
  \BibitemOpen
  \bibfield  {author} {\bibinfo {author} {\bibfnamefont {B.~J.}\ \bibnamefont
  {Carr}}, \bibinfo {author} {\bibfnamefont {K.}~\bibnamefont {Kohri}},
  \bibinfo {author} {\bibfnamefont {Y.}~\bibnamefont {Sendouda}}, \ and\
  \bibinfo {author} {\bibfnamefont {J.}~\bibnamefont {Yokoyama}},\ }\href
  {\doibase 10.1103/PhysRevD.81.104019} {\bibfield  {journal} {\bibinfo
  {journal} {Phys. Rev. D}\ }\textbf {\bibinfo {volume} {81}},\ \bibinfo
  {pages} {104019} (\bibinfo {year} {2010})},\ \Eprint
  {http://arxiv.org/abs/0912.5297} {arXiv:0912.5297 [astro-ph.CO]} \BibitemShut
  {NoStop}%
\bibitem [{\citenamefont {Laha}(2019)}]{Laha:2019ssq}%
  \BibitemOpen
  \bibfield  {author} {\bibinfo {author} {\bibfnamefont {R.}~\bibnamefont
  {Laha}},\ }\href {\doibase 10.1103/PhysRevLett.123.251101} {\bibfield
  {journal} {\bibinfo  {journal} {Phys. Rev. Lett.}\ }\textbf {\bibinfo
  {volume} {123}},\ \bibinfo {pages} {251101} (\bibinfo {year} {2019})},\
  \Eprint {http://arxiv.org/abs/1906.09994} {arXiv:1906.09994 [astro-ph.HE]}
  \BibitemShut {NoStop}%
\bibitem [{\citenamefont {Smyth}\ \emph {et~al.}(2020)\citenamefont {Smyth},
  \citenamefont {Profumo}, \citenamefont {English}, \citenamefont {Jeltema},
  \citenamefont {McKinnon},\ and\ \citenamefont
  {Guhathakurta}}]{Smyth:2019whb}%
  \BibitemOpen
  \bibfield  {author} {\bibinfo {author} {\bibfnamefont {N.}~\bibnamefont
  {Smyth}}, \bibinfo {author} {\bibfnamefont {S.}~\bibnamefont {Profumo}},
  \bibinfo {author} {\bibfnamefont {S.}~\bibnamefont {English}}, \bibinfo
  {author} {\bibfnamefont {T.}~\bibnamefont {Jeltema}}, \bibinfo {author}
  {\bibfnamefont {K.}~\bibnamefont {McKinnon}}, \ and\ \bibinfo {author}
  {\bibfnamefont {P.}~\bibnamefont {Guhathakurta}},\ }\href {\doibase
  10.1103/PhysRevD.101.063005} {\bibfield  {journal} {\bibinfo  {journal}
  {Phys. Rev. D}\ }\textbf {\bibinfo {volume} {101}},\ \bibinfo {pages}
  {063005} (\bibinfo {year} {2020})},\ \Eprint
  {http://arxiv.org/abs/1910.01285} {arXiv:1910.01285 [astro-ph.CO]}
  \BibitemShut {NoStop}%
\bibitem [{\citenamefont {Tisserand}\ \emph {et~al.}(2007)\citenamefont
  {Tisserand} \emph {et~al.}}]{EROS-2:2006ryy}%
  \BibitemOpen
  \bibfield  {author} {\bibinfo {author} {\bibfnamefont {P.}~\bibnamefont
  {Tisserand}} \emph {et~al.} (\bibinfo {collaboration} {EROS-2}),\ }\href
  {\doibase 10.1051/0004-6361:20066017} {\bibfield  {journal} {\bibinfo
  {journal} {Astron. Astrophys.}\ }\textbf {\bibinfo {volume} {469}},\ \bibinfo
  {pages} {387} (\bibinfo {year} {2007})},\ \Eprint
  {http://arxiv.org/abs/astro-ph/0607207} {arXiv:astro-ph/0607207} \BibitemShut
  {NoStop}%
\bibitem [{\citenamefont {Kavanagh}\ \emph {et~al.}(2018)\citenamefont
  {Kavanagh}, \citenamefont {Gaggero},\ and\ \citenamefont
  {Bertone}}]{Kavanagh:2018ggo}%
  \BibitemOpen
  \bibfield  {author} {\bibinfo {author} {\bibfnamefont {B.~J.}\ \bibnamefont
  {Kavanagh}}, \bibinfo {author} {\bibfnamefont {D.}~\bibnamefont {Gaggero}}, \
  and\ \bibinfo {author} {\bibfnamefont {G.}~\bibnamefont {Bertone}},\ }\href
  {\doibase 10.1103/PhysRevD.98.023536} {\bibfield  {journal} {\bibinfo
  {journal} {Phys. Rev. D}\ }\textbf {\bibinfo {volume} {98}},\ \bibinfo
  {pages} {023536} (\bibinfo {year} {2018})},\ \Eprint
  {http://arxiv.org/abs/1805.09034} {arXiv:1805.09034 [astro-ph.CO]}
  \BibitemShut {NoStop}%
\bibitem [{\citenamefont {Abbott}\ \emph {et~al.}(2019)\citenamefont {Abbott}
  \emph {et~al.}}]{LIGOScientific:2019kan}%
  \BibitemOpen
  \bibfield  {author} {\bibinfo {author} {\bibfnamefont {B.~P.}\ \bibnamefont
  {Abbott}} \emph {et~al.} (\bibinfo {collaboration} {LIGO Scientific,
  Virgo}),\ }\href {\doibase 10.1103/PhysRevLett.123.161102} {\bibfield
  {journal} {\bibinfo  {journal} {Phys. Rev. Lett.}\ }\textbf {\bibinfo
  {volume} {123}},\ \bibinfo {pages} {161102} (\bibinfo {year} {2019})},\
  \Eprint {http://arxiv.org/abs/1904.08976} {arXiv:1904.08976 [astro-ph.CO]}
  \BibitemShut {NoStop}%
\bibitem [{\citenamefont {Chen}\ and\ \citenamefont
  {Huang}(2020)}]{Chen:2019irf}%
  \BibitemOpen
  \bibfield  {author} {\bibinfo {author} {\bibfnamefont {Z.-C.}\ \bibnamefont
  {Chen}}\ and\ \bibinfo {author} {\bibfnamefont {Q.-G.}\ \bibnamefont
  {Huang}},\ }\href {\doibase 10.1088/1475-7516/2020/08/039} {\bibfield
  {journal} {\bibinfo  {journal} {JCAP}\ }\textbf {\bibinfo {volume} {08}},\
  \bibinfo {pages} {039} (\bibinfo {year} {2020})},\ \Eprint
  {http://arxiv.org/abs/1904.02396} {arXiv:1904.02396 [astro-ph.CO]}
  \BibitemShut {NoStop}%
\bibitem [{\citenamefont {Serpico}\ \emph {et~al.}(2020)\citenamefont
  {Serpico}, \citenamefont {Poulin}, \citenamefont {Inman},\ and\ \citenamefont
  {Kohri}}]{Serpico:2020ehh}%
  \BibitemOpen
  \bibfield  {author} {\bibinfo {author} {\bibfnamefont {P.~D.}\ \bibnamefont
  {Serpico}}, \bibinfo {author} {\bibfnamefont {V.}~\bibnamefont {Poulin}},
  \bibinfo {author} {\bibfnamefont {D.}~\bibnamefont {Inman}}, \ and\ \bibinfo
  {author} {\bibfnamefont {K.}~\bibnamefont {Kohri}},\ }\href {\doibase
  10.1103/PhysRevResearch.2.023204} {\bibfield  {journal} {\bibinfo  {journal}
  {Phys. Rev. Res.}\ }\textbf {\bibinfo {volume} {2}},\ \bibinfo {pages}
  {023204} (\bibinfo {year} {2020})},\ \Eprint
  {http://arxiv.org/abs/2002.10771} {arXiv:2002.10771 [astro-ph.CO]}
  \BibitemShut {NoStop}%
\bibitem [{\citenamefont {Hektor}\ \emph {et~al.}(2018)\citenamefont {Hektor},
  \citenamefont {H\"utsi}, \citenamefont {Marzola}, \citenamefont {Raidal},
  \citenamefont {Vaskonen},\ and\ \citenamefont {Veerm\"ae}}]{Hektor:2018qqw}%
  \BibitemOpen
  \bibfield  {author} {\bibinfo {author} {\bibfnamefont {A.}~\bibnamefont
  {Hektor}}, \bibinfo {author} {\bibfnamefont {G.}~\bibnamefont {H\"utsi}},
  \bibinfo {author} {\bibfnamefont {L.}~\bibnamefont {Marzola}}, \bibinfo
  {author} {\bibfnamefont {M.}~\bibnamefont {Raidal}}, \bibinfo {author}
  {\bibfnamefont {V.}~\bibnamefont {Vaskonen}}, \ and\ \bibinfo {author}
  {\bibfnamefont {H.}~\bibnamefont {Veerm\"ae}},\ }\href {\doibase
  10.1103/PhysRevD.98.023503} {\bibfield  {journal} {\bibinfo  {journal} {Phys.
  Rev. D}\ }\textbf {\bibinfo {volume} {98}},\ \bibinfo {pages} {023503}
  (\bibinfo {year} {2018})},\ \Eprint {http://arxiv.org/abs/1803.09697}
  {arXiv:1803.09697 [astro-ph.CO]} \BibitemShut {NoStop}%
\bibitem [{\citenamefont {Manshanden}\ \emph {et~al.}(2019)\citenamefont
  {Manshanden}, \citenamefont {Gaggero}, \citenamefont {Bertone}, \citenamefont
  {Connors},\ and\ \citenamefont {Ricotti}}]{Manshanden:2018tze}%
  \BibitemOpen
  \bibfield  {author} {\bibinfo {author} {\bibfnamefont {J.}~\bibnamefont
  {Manshanden}}, \bibinfo {author} {\bibfnamefont {D.}~\bibnamefont {Gaggero}},
  \bibinfo {author} {\bibfnamefont {G.}~\bibnamefont {Bertone}}, \bibinfo
  {author} {\bibfnamefont {R.~M.~T.}\ \bibnamefont {Connors}}, \ and\ \bibinfo
  {author} {\bibfnamefont {M.}~\bibnamefont {Ricotti}},\ }\href {\doibase
  10.1088/1475-7516/2019/06/026} {\bibfield  {journal} {\bibinfo  {journal}
  {JCAP}\ }\textbf {\bibinfo {volume} {06}},\ \bibinfo {pages} {026} (\bibinfo
  {year} {2019})},\ \Eprint {http://arxiv.org/abs/1812.07967} {arXiv:1812.07967
  [astro-ph.HE]} \BibitemShut {NoStop}%
\bibitem [{\citenamefont {Lu}\ \emph {et~al.}(2021)\citenamefont {Lu},
  \citenamefont {Takhistov}, \citenamefont {Gelmini}, \citenamefont {Hayashi},
  \citenamefont {Inoue},\ and\ \citenamefont {Kusenko}}]{Lu:2020bmd}%
  \BibitemOpen
  \bibfield  {author} {\bibinfo {author} {\bibfnamefont {P.}~\bibnamefont
  {Lu}}, \bibinfo {author} {\bibfnamefont {V.}~\bibnamefont {Takhistov}},
  \bibinfo {author} {\bibfnamefont {G.~B.}\ \bibnamefont {Gelmini}}, \bibinfo
  {author} {\bibfnamefont {K.}~\bibnamefont {Hayashi}}, \bibinfo {author}
  {\bibfnamefont {Y.}~\bibnamefont {Inoue}}, \ and\ \bibinfo {author}
  {\bibfnamefont {A.}~\bibnamefont {Kusenko}},\ }\href {\doibase
  10.3847/2041-8213/abdcb6} {\bibfield  {journal} {\bibinfo  {journal}
  {Astrophys. J. Lett.}\ }\textbf {\bibinfo {volume} {908}},\ \bibinfo {pages}
  {L23} (\bibinfo {year} {2021})},\ \Eprint {http://arxiv.org/abs/2007.02213}
  {arXiv:2007.02213 [astro-ph.CO]} \BibitemShut {NoStop}%
\bibitem [{\citenamefont {Carr}\ \emph {et~al.}(2020)\citenamefont {Carr},
  \citenamefont {Kohri}, \citenamefont {Sendouda},\ and\ \citenamefont
  {Yokoyama}}]{Carr:2020gox}%
  \BibitemOpen
  \bibfield  {author} {\bibinfo {author} {\bibfnamefont {B.}~\bibnamefont
  {Carr}}, \bibinfo {author} {\bibfnamefont {K.}~\bibnamefont {Kohri}},
  \bibinfo {author} {\bibfnamefont {Y.}~\bibnamefont {Sendouda}}, \ and\
  \bibinfo {author} {\bibfnamefont {J.}~\bibnamefont {Yokoyama}},\ }\href@noop
  {} {\  (\bibinfo {year} {2020})},\ \Eprint {http://arxiv.org/abs/2002.12778}
  {arXiv:2002.12778 [astro-ph.CO]} \BibitemShut {NoStop}%
\bibitem [{\citenamefont {Green}\ and\ \citenamefont
  {Kavanagh}(2021)}]{Green:2020jor}%
  \BibitemOpen
  \bibfield  {author} {\bibinfo {author} {\bibfnamefont {A.~M.}\ \bibnamefont
  {Green}}\ and\ \bibinfo {author} {\bibfnamefont {B.~J.}\ \bibnamefont
  {Kavanagh}},\ }\href {\doibase 10.1088/1361-6471/abc534} {\bibfield
  {journal} {\bibinfo  {journal} {J. Phys. G}\ }\textbf {\bibinfo {volume}
  {48}},\ \bibinfo {pages} {043001} (\bibinfo {year} {2021})},\ \Eprint
  {http://arxiv.org/abs/2007.10722} {arXiv:2007.10722 [astro-ph.CO]}
  \BibitemShut {NoStop}%
\end{thebibliography}%

\end{document}